\documentclass[floatfix,aps,amsmath,prd,twocolumn,eqsecnum,showpacs,nofootinbib,reprint]{revtex4-1}
\usepackage{hyperref,url}
\usepackage[hyphenbreaks]{breakurl}
\usepackage{graphicx,graphics,amsfonts,amsmath,amssymb,bm,psfrag,natbib,dcolumn,textcase}

\allowdisplaybreaks

\newcommand{\be}{\begin{equation}}
\newcommand{\ee}{\end{equation}}

\newcommand{\bs}{\begin{subequations}}
\newcommand{\es}{\end{subequations}}

\DeclareMathOperator{\csch}{csch}

\begin{document}

\title{The gravitational-wave memory from eccentric binaries}
\author{Marc Favata}
\thanks{NASA Postdoctoral Fellow}
\email{favata@tapir.caltech.edu}
\affiliation{Jet Propulsion Laboratory, 4800 Oak Grove Drive, Pasadena, CA 91109, USA}
\thanks{Copyright 2011 California Institute of Technology. Government sponsorship acknowledged.}
\affiliation{Theoretical Astrophysics, 350-17, California Institute of Technology, Pasadena, CA 91125, USA}
\date{Received 15 August 2011; published 6 December 2011}
\begin{abstract}
The nonlinear gravitational-wave memory causes a time-varying but nonoscillatory correction to the gravitational-wave polarizations. It arises from gravitational waves that are sourced by gravitational waves. Previous considerations of the nonlinear memory effect have focused on quasicircular binaries. Here, I consider the nonlinear memory from Newtonian orbits with arbitrary eccentricity. Expressions for the waveform polarizations and spin-weighted spherical-harmonic modes are derived for elliptic, hyperbolic, parabolic, and radial orbits. In the hyperbolic, parabolic, and radial cases the nonlinear memory provides a 2.5 post-Newtonian (PN) correction to the leading-order waveforms. This is in contrast to the elliptical and quasicircular cases, where the nonlinear memory corrects the waveform at \emph{leading} (0PN) order. This difference in PN order arises from the fact that the memory builds up over a short ``scattering'' time scale in the hyperbolic case, as opposed to a much longer radiation-reaction time scale in the elliptical case. The nonlinear memory corrections presented here
complete our knowledge of the leading-order (Peters--Mathews) waveforms for elliptical orbits. These calculations are also relevant for binaries with quasicircular orbits in the present epoch which had, in the past, large eccentricities. Because the nonlinear memory depends sensitively on the past evolution of a binary, I discuss the effect of this early-time eccentricity on the value of the late-time memory in nearly circularized binaries. I also discuss the observability of large ``memory jumps'' in a binary's past that could arise from its formation in a capture process. Lastly, I provide estimates of the signal-to-noise ratio of the linear and nonlinear memories from hyperbolic and parabolic binaries.
\end{abstract}
\pacs{04.25.Nx, 04.25.-g, 04.30.-w, 04.30.Db}
\maketitle
\section{\label{sec:intro}Introduction}
Gravitational waves (GWs) are usually thought of as purely oscillatory phenomena. For example, the GWs from the coalescence of compact-object binaries tend to have a characteristic structure: as a GW passes through a detector, the frequency and amplitude increase, but at late times, the amplitude exponentially decays to zero from its peak value. However, the GWs from a variety of sources display nonoscillatory components as well. The simplest example is the GWs produced by the scattering of two unbound masses in a hyperbolic orbit \cite{turner-unbound}. Other examples include the GW signal from the asymmetric ejection of matter \cite{burrows-hayesPRL96,kotake-sato-SNmemory,ott-murphy-burrows-2009} or neutrinos \cite{epstein-neutrinomemory,turner-neutrinomemory} in supernova explosions or gamma-ray-bursts \cite{sago-GRBmemory,segalis-ori-GRBmemory}. The GW signals from these sources all exhibit a property called GW \emph{memory}. This refers to a long time scale difference in the values of the observed metric perturbation associated with the GW:
\be
\label{eq:memdiff}
\Delta h_{+,\times} = \lim_{t\rightarrow +\infty} h_{+,\times} - \lim_{t\rightarrow -\infty} h_{+,\times},
\ee
where $h_{+,\times}$ are the two GW polarizations. In a GW detector that is truly freely-falling (one that follows geodesics), a GW with memory can cause a permanent deformation in the detector (hence the term \emph{memory}).

The nonoscillatory sources mentioned above are all examples of the \emph{linear memory} \cite{zeldovich-polnarev,braginskii-grishchuk,braginskii-thorne}. In these cases, the nonoscillatory component to the GW arises from nonoscillatory motions of the source that are encoded in the matter stress-energy tensor $T_{\mu \nu}$. Because unbound gravitating systems have sources that undergo nonoscillatory motions, the linear memory tends to occur in such systems. In addition to the linear memory, there is also a \emph{nonlinear memory} effect \cite{payne-zfl,blanchet-thesis,blanchet-damour-hereditary,christodoulou-mem}. This nonlinear memory arises from the gravitational waves produced by gravitational waves: it is sourced not by nonoscillatory motions encoded in $T_{\mu \nu}$ but rather in $T_{\mu \nu}^{\rm gw}$ \cite{isaacson,mtw}, which describes the stress-energy of radiated GWs. These radiated GWs are themselves always ``unbound'' and hence produce a memory \cite{kipmemory}. Furthermore, since this effect originates directly from the radiated GWs and not the motion of the source, the nonlinear memory is present in \emph{all} sources of GWs, including bounded systems.

Why study the nonlinear memory? One of the key goals of gravitational-wave astronomy is to experimentally probe our understanding of general relativity (GR). While solar system and binary pulsar tests can only probe GR in the weak-field regime where the theory is nearly Newtonian, observations of coalescing compact-object binaries (and especially merging binary black holes) will produce GWs whose properties will depend on the strong-field, highly-dynamical sector of GR. There are a variety of nonlinear effects which will imprint themselves on the waveforms from such systems. Of these effects the nonlinear memory is unique because (i) it describes waves produced by waves (among the most nonlinear of interactions present in the theory), (ii) its nonoscillatory imprint on the GW signal is distinct from the manifestation of other nonlinear effects, and (iii) although it arises from a higher-order nonlinear interaction, it can enter the post-Newtonian (PN) expansion of the waveform amplitude at leading (0PN) order.

The nonlinear memory also has the property of being \emph{hereditary}. This means that the nonlinear memory piece of the waveform amplitude at some observer's time $t$ depends not only on the configuration of the source at the corresponding retarded time $t-R$ (where $R$ is the distance to the source), but rather on the entire past-history of the source. To illustrate this property, compare the leading-order ``quadrupole'' expression for the GW field,
\be
\label{eq:hjk-quad}
h_{jk}^{\rm TT} = \frac{2}{R} \ddot{I}_{jk}^{\rm TT}(T_R),
\ee
with the corresponding expression for the nonlinear memory \cite{wiseman-will-memory},
\be
\label{eq:hjk-mem}
h_{jk}^{\rm TT\,(mem)} = \frac{4}{R} \int_{-\infty}^{T_R} dt' \left[ \int \frac{dE^{\rm gw}}{dt' d\Omega'} \frac{n'_j n'_k}{(1-{\bm n}' \cdot {\bm N})} d\Omega' \right]^{\rm TT},
\ee
where $I_{jk}$ is the mass quadrupole moment, $\frac{dE^{\rm gw}}{dt d\Omega}$ is the GW energy flux, $n_j$ is a unit radial vector, ${\bm N}$ is a unit vector pointing from the source to the observer, and ${\rm TT}$ means to take the transverse-traceless projection (see also Sec.~I of \cite{favata-pnmemory}). In Eq.~\eqref{eq:hjk-quad}, one can clearly see that the primary GWs measured by a detector are directly determined by the retarded time configuration of the source (as encoded by its mass quadrupole). But in the expression for the nonlinear memory [Eq.~\eqref{eq:hjk-mem}], its value at time $T_R$ depends on a integral into the infinite past.\footnote{The GW tail effect exhibits a similar property, but in that case, the integrand drops off more steeply in the past than does the memory, so only the nearby past need be considered (see Sec.~4 of \cite{arun25PNamp,*arun25PNampE} or Sec.~5.3.4 of \cite{maggiore-GWvol1} for a more detailed discussion).} So to determine the value of the nonlinear memory at one instant, one needs to know the energy flux (and hence the motion of the source) at all previous times.
\subsection{\label{subsec:motivation}Motivation}
Previous calculations of the nonlinear memory have focused almost exclusively on quasicircular binaries.
Early work focused on computing the memory during the inspiral phase for quasicircular orbits, starting with Wiseman and Will \cite{wiseman-will-memory}\footnote{Note that the expression for the memory in the text after Eq.~(17) of \cite{wiseman-will-memory} is missing a factor of $2$, but the curves in their Fig.~1 agree with the expressions below.} (see also \cite{kennefick-memory}). Remarkably, they found that the nonlinear memory affects the GW polarizations at leading (0PN) order:
\bs
\begin{align}
\label{eq:hplus-intro}
h_{+} &= -2\frac{\eta M}{R} x \left[ (1+c_{\Theta}^2) \cos2(\varphi - \Phi) - \frac{s^2_{\Theta}}{96}(17 + c_{\Theta}^2) \right],\\
\label{eq:hcross-intro}
h_{\times} &= -4\frac{\eta M}{R} x c_{\Theta} \sin2(\varphi - \Phi),
\end{align}
\es
where $M=m_1+m_2$ is the binary's total mass, $\eta=m_1 m_2/M^2$ is its reduced mass ratio, $x\equiv (M\omega)^{2/3}$ is the standard PN expansion variable for quasicircular binaries, $\varphi(t)$ is the orbital phase, $\omega=\dot{\varphi}$ is the orbital angular frequency, $c_{\Theta}\equiv \cos\Theta$, $s_{\Theta}\equiv \sin\Theta$, and $(\Theta,\Phi)$ are the angles in the source frame that point in the observer's direction ${\bm N}$. The second term in Eq.~\eqref{eq:hplus-intro} is the nonoscillatory memory term. (For a convenient and standard choice of the polarization triad, there is no nonlinear memory contribution to the $\times$ polarization.) In \cite{favata-pnmemory}, these nonlinear memory contributions were computed to 3PN order\footnote{Arun et.~al \cite{arun25PNamp,arun25PNampE} previously showed that the 0.5PN memory contribution vanishes.}, thus completing the PN expansion of the waveform amplitude consistently to that order.

More recent work has investigated in detail the nonlinear memory produced by merging binary black holes. Using a simple analytic model as well as an effective-one-body \cite{damour-nagar-EOBlecnotes2009} approach, the full evolution of the memory for the inspiral, merger, and ringdown was computed in \cite{favata-lisa7confproc,favata-memory-saturation} and the prospects for its detection with interferometers were examined.\footnote{Thorne \cite{kipmemory} also examined the memory's detectability, treating it as an unmodeled burst, while Kennefick \cite{kennefick-memory} considered only the inspiral contribution to the memory.} Calculations of the nonlinear memory from numerical relativity were performed in \cite{favata-PNNR-memory,pollney-reisswig-memory}, and its detectability with pulsar-timing-arrays was considered in \cite{seto-memory-MNRAS09,pshirkov-etal-memory,levin-vanHaasteren-memory}.

The purpose of this study is to analyze the behavior of the nonlinear memory for arbitrarily eccentric binaries. There are several reasons why this generalization is worth considering. First (and most importantly), real binaries will be eccentric and not quasicircular. Even though gravitational radiation-reaction tends to circularize binaries, we expect to observe some binaries with non-negligible ellipticity (see, e.g., Section I and Appendix A of \cite{yunes-arun-berti-will-eccentric-PRD2009} as well as \cite{wen-eccentricity-ApJ2003,oleary-etal-BHmergersGC-ApJ2006}), or with hyperbolic trajectories (from scattering events in clusters or galactic nuclei).
However, even if one is only interested in nearly circular binaries, the hereditary nature of the nonlinear memory makes it important to consider the eccentric case as well. Because binaries which are currently quasicircular were more eccentric in the past, this prior eccentricity has modified the orbital motion and could potentially influence the calculation of the nonlinear memory integral in Eq.~\eqref{eq:hjk-mem}. Clearly, if we hope to actually observe the nonlinear memory effect, we must also be prepared to account for binary eccentricity.

A second motivation comes simply from the desire to have a complete and consistent understanding of the waveforms produced in the general two-body problem. For example, waveform polarizations for elliptical binaries are implicit in the classic work by Peters and Mathews \cite{petersmathews} (although they focus on computing the radiated power), and are given explicitly first in Wahlquist \cite{wahlquist} and later in Refs.~\cite{moreno-garrido-etal-eccentric-MNRAS1994,moreno-etal-eccentric-MNRAS1995,pierro-pinto-etal-MNRAS2001} to leading (0PN) order. These elliptical waveforms have since been extended to 1PN order in amplitude by Junker and Sch\"{a}fer \cite{junker-schafer} (see also \cite{tessmergopamnras,majar-vasuth-unbound1pn-PRD2010}), to 1.5PN order in Blanchet and Sch\"{a}fer \cite{blanchetschafertails}, and to 2PN order (neglecting tails) in Gopakumar and Iyer \cite{GI2}.\footnote{Additional works also consider the GW polarizations in the case of eccentric binaries with spinning components, but for simplicity, I will not consider spin effects here. I will also not discuss results based on black hole perturbation theory, except to say that the nonlinear memory has not been considered in that formalism.} Frequency-domain waveforms are given in \cite{favata-phd,yunes-arun-berti-will-eccentric-PRD2009,tessmer-schafer-1pnfreqecc-PRD2010,tessmer-schafer-2pnfreqecc,favata-eccentricity-phasing}. (The GW \emph{phasing} for elliptical binaries is currently known to relative 3PN order in the conservative \cite{damourderuelle,damourschafer,schafer-wex-PLA1993,*schafer-wex-PLA1993-erratum,DGI,quasikep3PN,quasikepphasing35PN} and dissipative parts \cite{wagoner-will,*wagoner-will-erratum,blanchetschafer1PN,junker-schafer,blanchetschafertails,riethschafer,GI1,arun-eccentrictailsEflux,arun-eccentricEflux3PN,arun-etal-eccentric-orbitalelements-PRD2009,favata-phd,favata-eccentricity-phasing}.) As we will see later, even the leading-order (Peters--Mathews/Wahlquist) waveforms are incomplete because---as in the quasicircular case---the nonlinear memory modifies the polarizations at 0PN order.

In the case of hyperbolic orbits, waveform polarizations were first derived at 0PN order in Turner \cite{turner-unbound}, with 1PN corrections computed in \cite{junker-schafer,majar-vasuth-unbound1pn-PRD2010}. In the large-eccentricity (bremsstrahlung) limit, waveforms were computed to 1PN order in \cite{wagoner-will,*wagoner-will-erratum,turner-will} and using the ``post-linear'' formalism in \cite{kovacs-thorne-III,kovacs-thorne-IV}. These waveforms already show a linear memory, but the nonlinear memory has only been calculated in the large-eccentricity limit \cite{wiseman-will-memory}. Waveforms for radial orbits are also considered up to 1PN order in \cite{wagoner-will,*wagoner-will-erratum} (although the energy flux has been computed at 2PN \cite{simone-poisson-will-PRD1995} and 3PN \cite{iyer-mishra-radial3PNEflux-PRD2010} orders), but the nonlinear memory contribution in the radial case has not yet been computed.

In this paper, I will attempt to complete our knowledge of the nonlinear memory for binaries with arbitrary eccentricity, including elliptical, hyperbolic, parabolic and radial orbits. Unlike in Ref.~\cite{favata-pnmemory} where the nonlinear memory was computed to 3PN order in the quasicircular case, in this paper I will restrict the calculation to the leading-PN-order piece of the nonlinear memory.  I will also discuss how the nonlinear memory (because of its hereditary nature) is affected by the prior eccentricity of a nearly circularized binary.
\subsection{\label{subsec:summary}Summary}
In Sec.~\ref{subsec:general-mem}, we begin by reviewing the prescription in \cite{favata-pnmemory} for computing the nonlinear memory from the GW energy flux. This involves decomposing the GW polarizations into a sum of spin-weighted spherical harmonic modes $h_{lm}$, and relating the nonlinear memory modes $h_{lm}^{\rm (mem)}$ to ``lower-order'' oscillatory modes $h_{lm}^{\rm N}$ that are accurate to Newtonian (0PN) order. These Newtonian-order modes depend only on the mass quadrupole moment $I_{2m}$, for which explicit expressions are easily derived for general Newtonian binaries (Sec.~\ref{subsec:hNmodes}). In Sec.~\ref{subsubsec:kepler}, formulas for Keplerian orbits are reviewed, and general expressions for Keplerian waveforms are derived. The material in this section is concisely presented and may be useful to those interested in leading-order waveforms valid for any eccentricity.

In Sections \ref{subsec:elliptical}, \ref{subsec:hyperbolicorbits}, and \ref{subsec:radial}, I specialize the nonlinear memory calculation to the cases of elliptical, hyperbolic, parabolic, and radial orbits. The primary results are explicit expressions for the nonlinear memory modes $h_{lm}^{\rm (mem)}$ and the corresponding polarizations $h_{+,\times}$. Aside from these explicit expressions, the following are some of the primary results of this analysis: In the case of inspiralling elliptical binaries, the nonlinear memory behaves similarly to the quasicircular case. The primary contributions come from the $m=0$ modes, which have the scaling
\be
\label{eq:elliptc-scaling}
h_{l0}^{\rm (mem), ellip.} \propto \eta \frac{M}{R} \frac{M}{p} {\mathcal F}_{l0}(e_{-},e_t),
\ee
where $p(t)$ is the semilatus rectum of the ellipse and ${\mathcal F}_{lm}$ is a hypergeometric function that depends on the eccentricity $e_t(t)$ and its value $e_{-}$ at some early time [see Appendix \ref{app:hypergeom} for the exact expressions, or Eqs.~\eqref{eq:hlm-smallet} for the low-eccentricity limit]. Note that as in the quasicircular case, the nonlinear memory modifies the waveform at leading (Newtonian) order [cf.~Equation \eqref{eq:hplus-intro} with $M/p \sim O(x)$]. As discussed below, this result would hold even if we were to extend our analysis to orbits that undergo periastron advance. In the case of hyperbolic and parabolic orbits, this scaling is very different:
\be
\label{eq:hyper-scaling}
\Delta h_{lm}^{\rm (mem), hyperb.} \propto \eta^2 \frac{M}{R} \left(\frac{M}{p}\right)^{7/2} {\mathcal H}_{lm}(e_t),
\ee
where ${\mathcal H}_{lm}$ [ which can be read off of Eqs.~\eqref{eq:Deltahlm-hyperbolic}] is a function of the eccentricity $e_t$ (which is constant in the absence of radiation reaction) . Note that in this case all of the $(l,m)$ modes contribute to the nonlinear memory, which is a factor of $\eta (M/p)^{5/2}$ smaller than in the elliptical case. (A similar scaling also holds in the case of radial orbits.)

It is easy to understand the reason for this difference in scalings. The nonlinear memory can be written as a time integral of the form
\be
\label{eq:hmemintegral}
h_{lm}^{\rm (mem)} = \int_{-\infty}^{T_R} dt\, {h}_{lm}^{\rm (mem) (1)},
\ee
where the time derivative $h_{lm}^{\rm (mem)(1)}\equiv dh_{lm}^{\rm (mem)}/dt$ has the same leading-order scaling [$\propto \eta^2 (M/p)^5$] for all orbits. The difference in the scalings results from the time scale on which the integration is carried out. In the case of elliptical orbits, one must integrate over the entire inspiral, which occurs on a radiation-reaction time scale $T_{\rm rr}$, so that
\be
\Delta h_{lm}^{\rm (mem),\,ellip.} \sim h_{lm}^{\rm (mem)(1)} T_{\rm rr} \sim h_{lm}^{\rm (mem)(1)} \frac{M}{\eta} \left(\frac{p}{M}\right)^{4}.
\ee
In the hyperbolic case, one effectively integrates over a much shorter ``scattering'' or ``orbital'' time scale $T_{\rm orb}$,\footnote{To be precise, one is still integrating over the infinite time that the hyperbolic orbit spans, but the contribution to the memory integral mostly builds up over the short amount of time it takes the binary to ``scatter'' or significantly change its direction. This ``scattering time'' is not precisely defined, but it scales the same way (via Kepler's law) as what we would normally call an ``orbital time'' in the case of bound orbits.} so
\be
\Delta h_{lm}^{\rm (mem),\,hyperb.} \sim h_{lm}^{\rm (mem)(1)} T_{\rm orb} \sim h_{lm}^{\rm (mem)(1)} M \left(\frac{p}{M}\right)^{3/2}.
\ee
Since
\be
\frac{T_{\rm orb}}{T_{\rm rr}} \sim \eta \left( \frac{M}{p} \right)^{5/2},
\ee
we can see why the nonlinear memory in the hyperbolic/parabolic case enters at a much higher PN order than the elliptical/quasicircular case. Integrating over the much longer radiation-reaction time effectively allows the memory to ``build up'' to a much larger value than one would naively expect from such a high-order PN effect.

One of the more important results from this study concerns the analysis of the dependence of the memory on the early-time history of the binary (Sec.~\ref{sec:earlymemory}). For example, all quasicircular binaries presumably had elliptical orbits earlier in their evolution, and this increasing (into the past) ellipticity could eventually result in a hyperbolic binary. To address this issue, I computed the evolution of the dominant mode of the nonlinear memory waveform as a function of time for a nearly circular binary. This evolution was computed (i) assuming that the binary always remains quasicircular, and (ii) assuming that the binary's eccentricity increases as time evolves into the past. A comparison of these two evolutions is shown in Fig.~\ref{fig:h20-circ-ecc} (where time is parameterized in terms of the eccentricity of the elliptical binary). Accounting for the evolving eccentricity makes a small (but non-negligible) correction to the memory. The eccentricity correction is small because even though eccentricity is increasing into the past, so is the orbital separation (or the semilatus rectum). Since the integrand in Eq.~\eqref{eq:hmemintegral} is weighted by a factor of $(M/p)^5$, its value drops off at larger separations; so late-time values (smaller $p$, when $e_t$ is also small) are weighted more heavily than the distant past (when the eccentricity is large, but $M/p$ is very small).

An eccentric binary could also produce a large memory jump in its distant past, for example due to its sudden formation in a capture process. Such a memory jump would in principle be observable, but only if one's GW detector is operating when (in retarded time) that jump occurred. In general, effects of the early-time history of the binary are not observable in the memory signal if they occurred before the start of the observation. Hereditary effects are therefore only important over the observation time, and one need not worry about knowing the state of the binary prior to the start of the observation. This is illustrated with an explicit example in Sec.~\ref{sec:earlymemory} and Fig.~\ref{fig:deltaxmemory}.

Lastly, in Sec.~\ref{sec:estimates}, I discuss how to use the waveforms for hyperbolic obits to make rough estimates for the signal-to-noise ratio (SNR) of a memory signal from a gravitational-scattering event. Unlike the case of bound, inspiralling binaries (where the innermost stable circular orbit or the ringdown provides a natural high-frequency cutoff), the SNRs from hyperbolic binaries depend on a distance of closest approach (as well as an eccentricity parameter) and are dominated by the linear rather than the nonlinear memory. For future ground-based detectors, linear and nonlinear memory signals from the scattering of stellar-mass binaries within our galaxy are potentially detectable. Space-based detectors could see linear memory signals from the scattering of supermassive black holes (SMBHs), but detecting the nonlinear memory from these events will be more difficult. (This is in contrast to the nonlinear memory from \emph{merging} SMBHs, which should be detectable to moderate redshifts \cite{favata-memory-saturation,favata-lisa7confproc,favata-amaldiconfproc-memory-CQG2010,favata-PNNR-memory}.)

Some useful results are presented in the Appendices. Appendix \ref{app:moments} derives general expressions for the spherical harmonic modes of the mass and current source multipole moments at Newtonian order. Appendix \ref{app:hypergeom} shows how to express certain integrals over eccentricity in terms of hypergeometric functions. Appendix \ref{app:averaging} discusses the role of wavelength averaging of the GW stress-energy tensor in the nonlinear memory calculation.
\section{\label{sec:compute-mem}Computing the nonlinear memory}
We begin this section by first reviewing the decomposition of the GW polarizations in terms of multipole modes $h_{lm}$ on a spin-weighted spherical harmonic basis (Sec.~\ref{subsec:general-mem}). The nonlinear memory pieces of these modes [$h^{\rm (mem)}_{lm}$] are related to integrals over the GW energy flux $\frac{dE_{\rm gw}}{dtd\Omega}$. This is described in \cite{favata-pnmemory}, and the reader is directed there for a more detailed exposition. Since we are only interested in the leading-order memory, it is sufficient to express the energy flux in terms of the Newtonian-order nonmemory\footnote{When referring to ``nonmemory'' modes here and below, I really mean the ``non-nonlinear-memory'' pieces of the modes, which could contain oscillatory or linear memory pieces.} modes $h^{\rm N}_{lm}$. Section \ref{subsec:hNmodes} gives explicit expressions for these Newtonian-order modes and specializes them to Keplerian orbits parameterized in terms of the true anomaly angle $v$. These modes are then substituted into the expression for the nonlinear memory modes (Sec.~\ref{subsubsec:kepler}). Sections \ref{subsec:elliptical}, \ref{subsec:hyperbolicorbits}, and \ref{subsec:radial} specialize the calculation to the case of elliptical, hyperbolic, parabolic, and radial orbits, providing explicit expressions for the $h^{\rm (mem)}_{lm}$ modes and their corresponding $+$ and $\times$ polarizations.
\subsection{\label{subsec:general-mem}General expressions for the waveform and nonlinear-memory modes}
The GW polarizations can be decomposed as a sum over multipole modes via
\be
\label{eq:hdecompose}
h_{+} - i h_{\times} = \sum_{l=2}^{\infty} \sum_{m=-l}^{l} h^{lm} {}_{-2}Y^{lm}(\Theta,\Phi) \;,
\ee
where
\be
\label{eq:hlm}
h^{lm} = \frac{G}{\sqrt{2} R c^{l+2}} \left[ U^{lm}(T_R) - \frac{i}{c} V^{lm}(T_R) \right].
\ee
Here, ${}_{-2}Y^{lm}(\Theta,\Phi)$ are the spin-weighted spherical harmonics, $(R, \Theta, \Phi)$ are the distance and angles that point from the source to the observer, and $U^{lm}$ and $V^{lm}$ are the spherical harmonic representations of the radiative mass and current multipole moments (see Sec.~II A of \cite{favata-pnmemory} for details and notation). Constructing the GW polarizations requires explicit expressions for the ${}_{-2}Y^{lm}(\Theta,\Phi)$. The general formula for the  ${}_{-2}Y^{lm}(\Theta,\Phi)$ can be found in Eqs.~(2.13) and (2.14) of \cite{favata-pnmemory}. Here, we will only need the following modes:
\bs
\label{eq:m2Ylm}
\begin{align}
{}_{-2}Y^{20} &= \frac{3}{4} \sqrt{\frac{5}{6\pi}} {\rm s}^2_{\Theta},\\
{}_{-2}Y^{2\pm2} &= \frac{1}{8} \sqrt{\frac{5}{\pi}} (1\pm {\rm c}_{\Theta})^2 e^{\pm 2i\Phi},\\
{}_{-2}Y^{40} &= \frac{3}{8} \sqrt{\frac{5}{2\pi}} {\rm s}^2_{\Theta} (7 {\rm c}^2_{\Theta} -1),\\
{}_{-2}Y^{4\pm2} &= \frac{3}{8} \sqrt{\frac{1}{\pi}} (1\pm {\rm c}_{\Theta})^2 (1 \mp 7 {\rm c}_{\Theta} + 7 {\rm c}^2_{\Theta}) e^{\pm 2i\Phi},\\
{}_{-2}Y^{4\pm4} &= \frac{3}{16} \sqrt{\frac{7}{\pi}} {\rm s}^2_{\Theta} (1\pm {\rm c}_{\Theta})^2 e^{\pm 4i\Phi},
\end{align}
\es
where we define ${\rm c}_{\Theta}\equiv \cos\Theta$ and ${\rm s}_{\Theta}\equiv \sin\Theta$.

At leading order, the radiative moments $U_{lm}$ and $V_{lm}$ reduce to the source moments $I_{lm}$ and $J_{lm}$. At higher PN orders, the radiative moments are corrected by tail terms and other nonlinear couplings [see, e.g., Eqs.~(2.21), (2.22), and (2.32) of \cite{favata-pnmemory}]. Since we are focused on only the leading-order memory contribution, we will ignore all of these higher-order terms except for the nonlinear memory contribution $U_{lm}^{\rm (mem)}$ itself. Note that there is no nonlinear memory contribution to $V_{lm}$. For our purposes, we can therefore ignore all current multipole moments. This allows us to approximate the waveform modes as\footnote{We denote $n^{\rm th}$ time derivatives by $A^{(n)} \equiv d^nA/dt^n$.}
\be
\label{eq:hlmsimp}
h_{lm} \approx \frac{1}{\sqrt{2} R} I_{lm}^{(l)} + h_{lm}^{\rm (mem)}.
\ee
Here, the nonlinear memory piece is given by [see Eqs.~(2.32) and (3.3) of \cite{favata-pnmemory}]
\begin{multline}
\label{eq:hlmmem}
h_{lm}^{\rm (mem)} = \frac{16\pi}{R} \sqrt{\frac{(l-2)!}{(l+2)!}} \int_{-\infty}^{T_R} \!\! dt \int \! d\Omega \, \frac{dE_{\rm gw}}{dt d\Omega}(\Omega) Y_{lm}^{\ast}(\Omega) \\
= R \sqrt{\frac{(l-2)!}{(l+2)!}} \sum_{l'=2}^{\infty} \sum_{l''=2}^{\infty} \sum_{m'=-l'}^{l'} \sum_{m''=-l''}^{l''} (-1)^{m+m''}
\\
\times G^{2 -2 0}_{l' l'' l m' -m'' -m} \int_{-\infty}^{T_R} \!\! dt \left\langle \dot{h}_{l'm'} \dot{h}^{\ast}_{l''m''} \right\rangle  ,
\end{multline}
where in the second line we have substituted the expansion for the energy flux $\frac{dE_{\rm gw}}{dtd\Omega}$ in terms of the $h_{lm}$ modes [Eq.~(2.28) of \cite{favata-pnmemory}] and $G^{s_1 s_2 s_3}_{l_1 l_2 l_3 m_1 m_2 m_3}$ is an angular integral proportional to the product of three spin-weighted spherical harmonics (see Appendix A of \cite{favata-pnmemory}). The angle brackets $\langle \; \rangle$ mean to average over several wavelengths of the GW and arise from the averaging implicit in the construction of a well-defined GW stress-energy tensor \cite{mtw,isaacson} (see Appendix \ref{app:averaging} for a discussion of the implications of this averaging).

Note that the nonlinear memory modes $h_{lm}^{\rm (mem)}$ are themselves defined in terms of the full $h_{lm}$ modes. But in practice, the ``nonlinear memory contribution to the nonlinear memory'' is negligible, so we only need substitute the nonmemory modes into the right-hand-side of Eq.~\eqref{eq:hlmmem}. Furthermore, since we are only concerned with the leading-order nonlinear memory, we need only substitute the leading-PN-order piece of the $l=2$ mode $h_{2m}$ into the right-hand-side of Eq.~\eqref{eq:hlmmem}.

Now we proceed to compute the angular integrals in Eq.~\eqref{eq:hlmmem} and explicitly express the $h_{lm}^{\rm (mem)}$ in terms of the individual $h_{2m}$ modes. We begin by noting that the angular integration implies certain selection rules on the maximum $l$ for which the $h_{lm}^{\rm (mem)}$ are nonzero (see Sec.~III B of \cite{favata-pnmemory}). Since our calculation is to leading-order, only the $h_{2m}$ modes enter the right-hand-side of Eq.~\eqref{eq:hlmmem}. The selection rules then imply that only $h_{2m}^{\rm (mem)}$, $h_{3m}^{\rm (mem)}$, and $h_{4m}^{\rm (mem)}$ will be nonzero; all higher-$l$ nonlinear memory modes vanish (this was checked by explicit calculation).

Our results are further simplified by the fact that the nonmemory piece of the $h_{lm}$ modes are approximated by
\be
\label{eq:hlmN}
h^{\rm N}_{lm} \equiv \frac{I^{(l)}_{lm}}{R\sqrt{2}},
\ee
where the ${\rm N}$ emphasizes that our results are valid only at Newtonian order. Since $I_{lm}^{\ast} = (-1)^m I_{l -m}$, the nonmemory modes also satisfy  $h^{{\rm N}\,\ast}_{lm}= (-1)^m h^{\rm N}_{l -m}$. We also note that since $I_{lm} \propto Y_{lm}^{\ast}(\theta,\varphi)$ (see Appendix \ref{app:moments}), and if we assume that the orbit lies in the $x$--$y$ plane (so that $\theta=\pi/2$), then $h^{\rm N}_{2 \pm 1} \propto Y_{2 \pm 1}^{\ast} \propto \sin2\theta =0$. These simplifications imply
\be
h^{{\rm N}\,\ast}_{20} = h^{\rm N}_{20}, \;\; h^{{\rm N}\,\ast}_{2 \pm 2} = h^{\rm N}_{2 \mp 2}, \;\; h^{\rm N}_{2 \pm 1} = h^{{\rm N}\,\ast}_{2 \pm 1}=0.
\ee

Defining $h^{\rm (mem)(1)}_{lm}\equiv d h^{\rm (mem)}_{lm} /dT_R$, explicitly evaluating the angular integrals in Eq.~\eqref{eq:hlmmem} (with $\dot{h}_{lm} \rightarrow \dot{h}^{\rm N}_{lm}$ on the right-hand-side) then yields
\bs
\label{eq:hlmmem-hlm}
\begin{align}
h^{\rm (mem)(1)}_{2 \pm 1} &= h^{\rm (mem)(1)}_{3 m} = h^{\rm (mem)(1)}_{4 \pm 1} = h^{\rm (mem)(1)}_{4 \pm 3} = 0, \\
h^{\rm (mem)(1)}_{2 0} &= \frac{R}{42} \sqrt{\frac{15}{2\pi}} \left\langle 2 |\dot{h}^{\rm N}_{22}|^2 - |\dot{h}^{\rm N}_{20}|^2 \right\rangle, \\
h^{\rm (mem)(1)}_{2 \pm 2} &= \frac{R}{21} \sqrt{\frac{15}{2\pi}} \left\langle \dot{h}^{\rm N}_{20} \dot{h}^{\rm N}_{2 \pm 2} \right\rangle, \\
h^{\rm (mem)(1)}_{4 0} &= \frac{R}{1260} \sqrt{\frac{5}{2\pi}} \left\langle |\dot{h}^{\rm N}_{22}|^2 + 3 |\dot{h}^{\rm N}_{20}|^2 \right\rangle, \\
h^{\rm (mem)(1)}_{4 \pm 2} &= \frac{R}{252} \sqrt{\frac{3}{2\pi}} \left\langle \dot{h}^{\rm N}_{20} \dot{h}^{\rm N}_{2 \pm 2} \right\rangle , \\
h^{\rm (mem)(1)}_{4 \pm 4} &= \frac{R}{504} \sqrt{\frac{14}{2\pi}}  \left\langle (\dot{h}^{\rm N}_{2 \pm 2})^2 \right\rangle .
\end{align}
\es
\subsection{\label{subsec:hNmodes}Explicit expressions for the $h^{\rm N}_{lm}$ and $h^{\rm (mem)(1)}_{lm}$ modes for Newtonian binaries}
Now we write out explicit expressions for the $h^{\rm N}_{lm}$ modes. A derivation of the source mass and current multipole moments for Newtonian binaries is given in Appendix \ref{app:moments}. The result for the mass quadrupole moment is found in Eq.~\eqref{eq:Ilmxy}. The $l=2$ moments for an orbit in the $x$\mbox{--}$y$ plane are
\bs
\begin{align}
I^{\rm N}_{20} &= -4 \sqrt{\frac{\pi}{15}} \eta M r(t)^2, \\
I^{\rm N}_{2 \pm 2} &= 2 \sqrt{\frac{2 \pi}{5}} \eta M r(t)^2 e^{\mp 2 i \varphi(t)},
\end{align}
\es
where $M=m_1+m_2$, $\eta=m_1 m_2/M^2$, $r(t)$ is the relative orbital separation, and $\varphi(t)$ is the relative orbital phase.

Next we compute time derivatives of these mass moments. We eliminate second derivatives using the Newtonian equations of motion:
\bs
\label{eq:Newteqns}
\begin{align}
\ddot{r} &= r \dot{\varphi}^2 - \frac{M}{r^2} , \\
\ddot{\varphi} &= - \frac{2 \dot{r} \dot{\varphi}}{r}.
\end{align}
\es
Dropping the ``${\rm N}$'' label, the resulting derivatives of the moments are
\bs
\label{eq:dIlm}
\begin{align}
{I}_{20}^{(1)} &= -8 \sqrt{\frac{\pi}{15}} \eta M r \dot{r},  \\
{I}_{2 \pm 2}^{(1)} &= 4 \sqrt{\frac{2 \pi}{5}} \eta M r \left( \dot{r} \mp i r \dot{\varphi} \right) e^{\mp 2 i \varphi} , \\
{I}_{20}^{(2)} &= -8 \sqrt{\frac{\pi}{15}} \eta M \left( \dot{r}^2 + r^2 \dot{\varphi}^2 - \frac{M}{r} \right) , \\
{I}_{2 \pm 2}^{(2)} &= 4 \sqrt{\frac{2\pi}{5}} \eta M e^{\mp 2 i \varphi} \left( \dot{r}^2 - r^2 \dot{\varphi}^2 - \frac{M}{r} \mp 2 i r \dot{r} \dot{\varphi} \right) , \\
{I}_{20}^{(3)} &= 8 \sqrt{\frac{\pi}{15}} \eta \left(\frac{M}{r} \right)^2 \dot{r} , \\
{I}_{2 \pm 2}^{(3)} &= -4 \sqrt{\frac{2\pi}{5}} \eta \left( \frac{M}{r} \right)^2 e^{\mp 2 i \varphi} \left( \dot{r} \mp 4 i r \dot{\varphi} \right).
\end{align}
\es
These expressions are valid for general orbits that satisfy the Newtonian equations of motion \eqref{eq:Newteqns}. The explicit Newtonian-order polarizations are found from substituting Eqs.~\eqref{eq:hlmN} and \eqref{eq:m2Ylm} into Eq.~\eqref{eq:hdecompose} and summing only over the $l=2$ modes:\footnote{These formulas agree with Eqs.~(6) of \cite{DGI} if we choose $\Phi=\pm\pi/2$ and change the overall sign on both polarizations. This difference arises from the choice of the polarization tensors (see Sec.~IIA of \cite{favata-pnmemory} for the convention used here).}
\bs
\label{eq:hpx}
\begin{multline}
h_{+}^{\rm N} = \frac{\eta M}{R} \left\{ (1+{\rm c}^2_{\Theta})\left[ \left(\dot{r}^2 - r^2 \dot{\varphi}^2 -\frac{M}{r} \right) \cos2(\varphi-\Phi)
\right. \right. \\ \left. \left.
- 2 \dot{r} r \dot{\varphi} \sin2(\varphi-\Phi) \right]
- {\rm s}^2_{\Theta} \left( \dot{r}^2 + r^2 \dot{\varphi}^2 - \frac{M}{r} \right) \right\},
\end{multline}
\begin{multline}
h_{\times}^{\rm N} = 2\frac{\eta M}{R} {\rm c}_{\Theta} \bigg[ 2 \dot{r} r \dot{\varphi} \cos2(\varphi-\Phi)
\\
+ \left( \dot{r}^2 - r^2 \dot{\varphi}^2 - \frac{M}{r} \right) \sin2(\varphi-\Phi) \bigg].
\end{multline}
\es

\begin{figure*}[t]
$
\begin{array}{ccc}
\includegraphics[angle=0, width=0.44\textwidth]{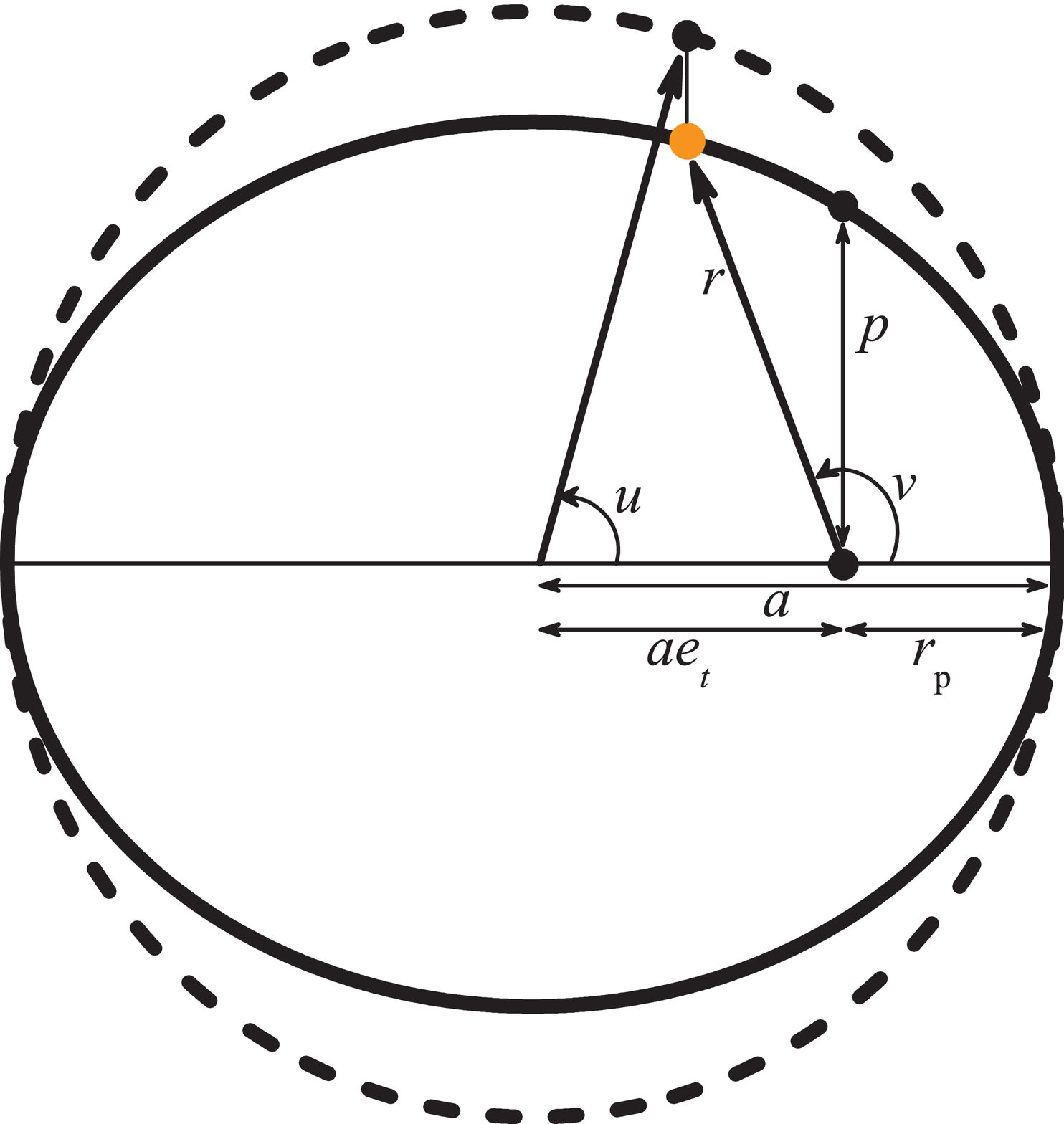} & \;\;\;\;\;\;\;\;\;\;\;\;\;\;\; &
\includegraphics[angle=0, width=0.4\textwidth]{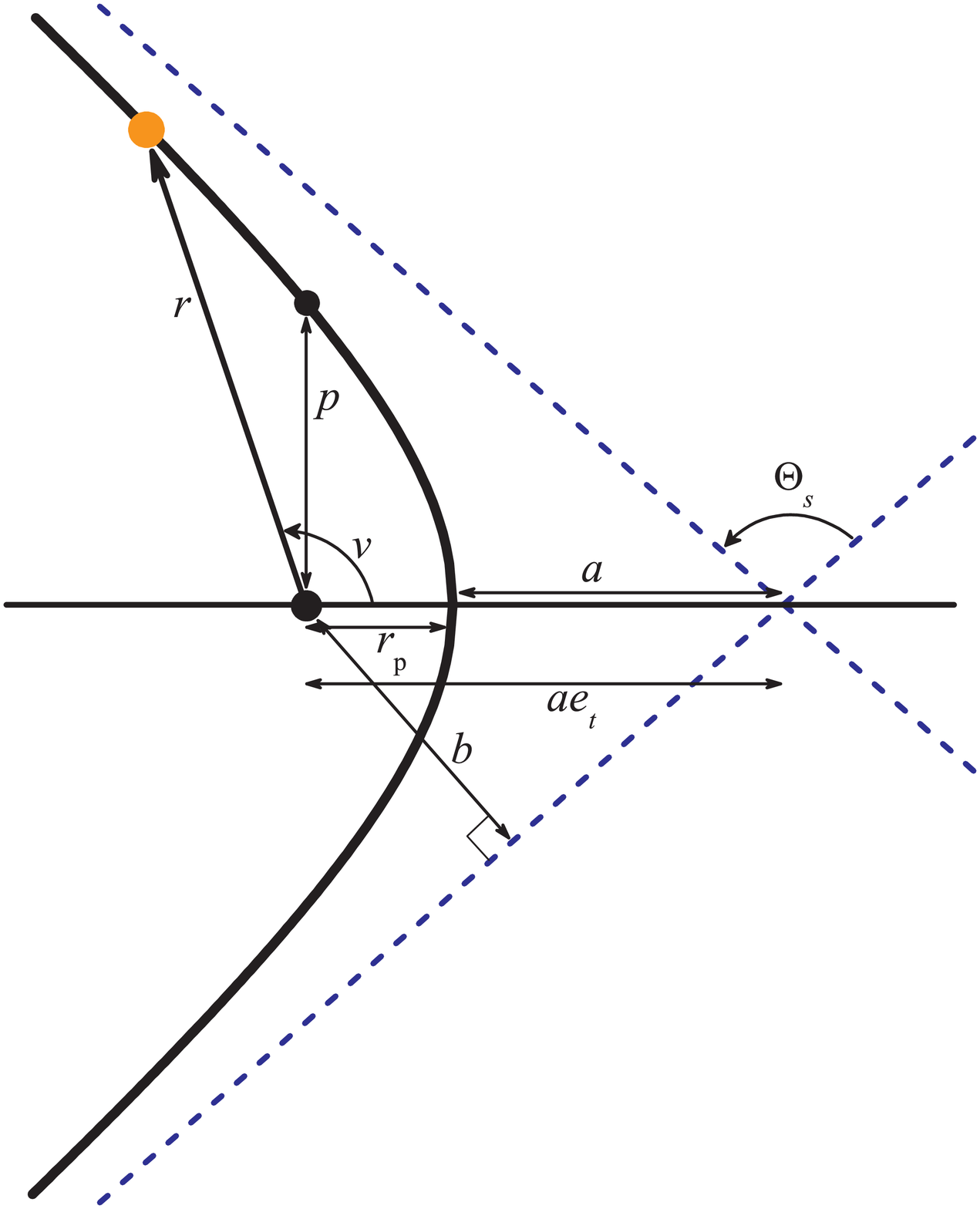}
\end{array}
$
\caption{\label{fig:diagrams}(color online). Notation and parameters describing elliptical and hyperbolic orbits. The left-hand figure shows a particle moving on an ellipse at a distance $r$ from the focus and making an angle $v$ (the true anomaly) with respect to the pericenter. The ellipse has an eccentricity $e_t$ and a size described by either its semimajor axis $a$, focus-pericenter distance $r_{\rm p}$, or semilatus rectum $p$. Also indicated is the eccentric anomaly $u$, which is the angle from the pericenter to the projection of the particle's position on the circle that circumscribes the ellipse. The right-hand figure shows a particle moving on a hyperbolic orbit, where we additionally indicate the asymptotes of the hyperbola (dashed lines), the particle's impact parameter $b$, and the scattering angle $\Theta_s$. The argument of pericenter $\varpi$ is taken to be zero in these diagrams.}
\end{figure*}
Substituting $\dot{h}^{\rm N}_{2m} = I^{(3)}_{2m}/(R \sqrt{2})$ and Eqs.~\eqref{eq:dIlm} into Eqs.~\eqref{eq:hlmmem-hlm} then gives
\bs
\label{eq:dhmem-newt}
\begin{align}
h^{\rm (mem)(1)}_{20} & = \frac{16}{21 R} \sqrt{\frac{2\pi}{15}} \eta^2 \left\langle \left( \frac{M}{r} \right)^4 \left( \dot{r}^2 + 24 r^2 \dot{\varphi}^2 \right) \right\rangle   , \\
h^{\rm (mem)(1)}_{2 \pm 2} & = -\frac{16}{21 R} \sqrt{\frac{\pi}{5}} \eta^2 \left\langle \left( \frac{M}{r} \right)^4  \dot{r} e^{\mp 2 i \varphi} \left( \dot{r} \mp 4 i r \dot{\varphi} \right) \right\rangle  , \\
h^{\rm (mem)(1)}_{40} & =   \frac{2}{315 R} \sqrt{\frac{2\pi}{5}} \eta^2 \left\langle \left( \frac{M}{r} \right)^4 \left( 3 \dot{r}^2 + 16 r^2 \dot{\varphi}^2 \right) \right\rangle   , \\
h^{\rm (mem)(1)}_{4 \pm 2} & =  -\frac{4}{315 R} \sqrt{\pi} \eta^2 \left\langle \left( \frac{M}{r} \right)^4 \dot{r} e^{\mp 2 i \varphi}  \left( \dot{r} \mp 4 i r \dot{\varphi} \right) \right\rangle  , \\
h^{\rm (mem)(1)}_{4 \pm 4} & = \frac{2}{45 R} \sqrt{\frac{\pi}{7}} \eta^2 \left\langle \left( \frac{M}{r} \right)^4 e^{\mp 4 i \varphi} \left( \dot{r} \mp 4 i r \dot{\varphi} \right)^2 \right\rangle ,
\end{align}
\es
for the nonvanishing memory modes.
\subsubsection{\label{subsubsec:kepler}Formulas for Keplerian orbits}
Now we wish to specialize these expressions to Keplerian orbits. The time evolution of the orbital separation $r$ and phase $\varphi$ is parameterized in terms of the true anomaly $v$ (not to be confused with the orbital speed $V$),
\bs
\label{eq:keperlian}
\begin{align}
r &= \frac{p}{1+e_t \cos v} , \\
v &= \varphi - \varpi , \\
\label{eq:dphidt}
\dot{v} &= \dot{\varphi} = \frac{\sqrt{p M}}{r^2},
\end{align}
\es
where, for planar orbits, only three orbital elements are needed to parameterize the binary: the semilatus rectum $p$, the eccentricity\footnote{Throughout this article, we denote the eccentricity by $e_t$ to avoid confusion with the mathematical constant $e$ and to emphasize that the eccentricity can evolve with time [ie., $e_t=e_t(t)$]. This choice is not meant to imply an identification of our eccentricity parameter with the ``time eccentricity'' used in the quasi-Keplerian formalism that describes PN elliptical orbits (e.g., \cite{DGI} and references therein). In that formalism, three eccentricity parameters, $e_t$, $e_r$, and $e_{\varphi}$, are introduced; but at Newtonian order these three eccentricities are equivalent and we can identify either of them with the $e_t$ used here.} $e_t$, and the argument of pericenter $\varpi$. Figure \ref{fig:diagrams} illustrates the meaning of the various orbital parameters introduced throughout this article. Time is determined by integrating Eq.~\eqref{eq:dphidt}:
\be
\label{eq:time-eqn}
t - t_0 = \sqrt{\frac{p^3}{M}} \int \frac{dv}{(1+e_t \cos v)^2} .
\ee
The following derivatives follow from Eqs.~\eqref{eq:keperlian}:
\bs
\label{eq:kepderivatives}
\begin{align}
\dot{r} &= e_t \sqrt{\frac{M}{p}} \sin v , \\
\ddot{r} &= \frac{M}{r^3} (p - r) , \\
\ddot{\varphi} &= - \frac{2M}{r^3} e_t \sin v.
\end{align}
\es
Note also that the instantaneous orbital velocity is given by
\be
V^2=\dot{r}^2+r^2\dot{\varphi}^2 = \frac{M}{p} (1+e_t^2 + 2e_t \cos v),
\ee
the orbital energy per reduced mass is
\be
\frac{E}{\mu} = \frac{V^2}{2}-\frac{M}{r} = -\frac{1}{2} \frac{M}{p} (1-e_t^2),
\ee
and the semilatus rectum is related to the semi-major axis $a$ by
\be
\label{eq:semilatusdefine}
p = a |1-e_t^2|
\ee
and to the pericenter distance $r_{\rm p}$ by
\be
\label{eq:rperi}
p=r_{\rm p} (1+e_t).
\ee
\begin{figure*}[t]
$
\begin{array}{cc}
\includegraphics[angle=0, width=0.48\textwidth]{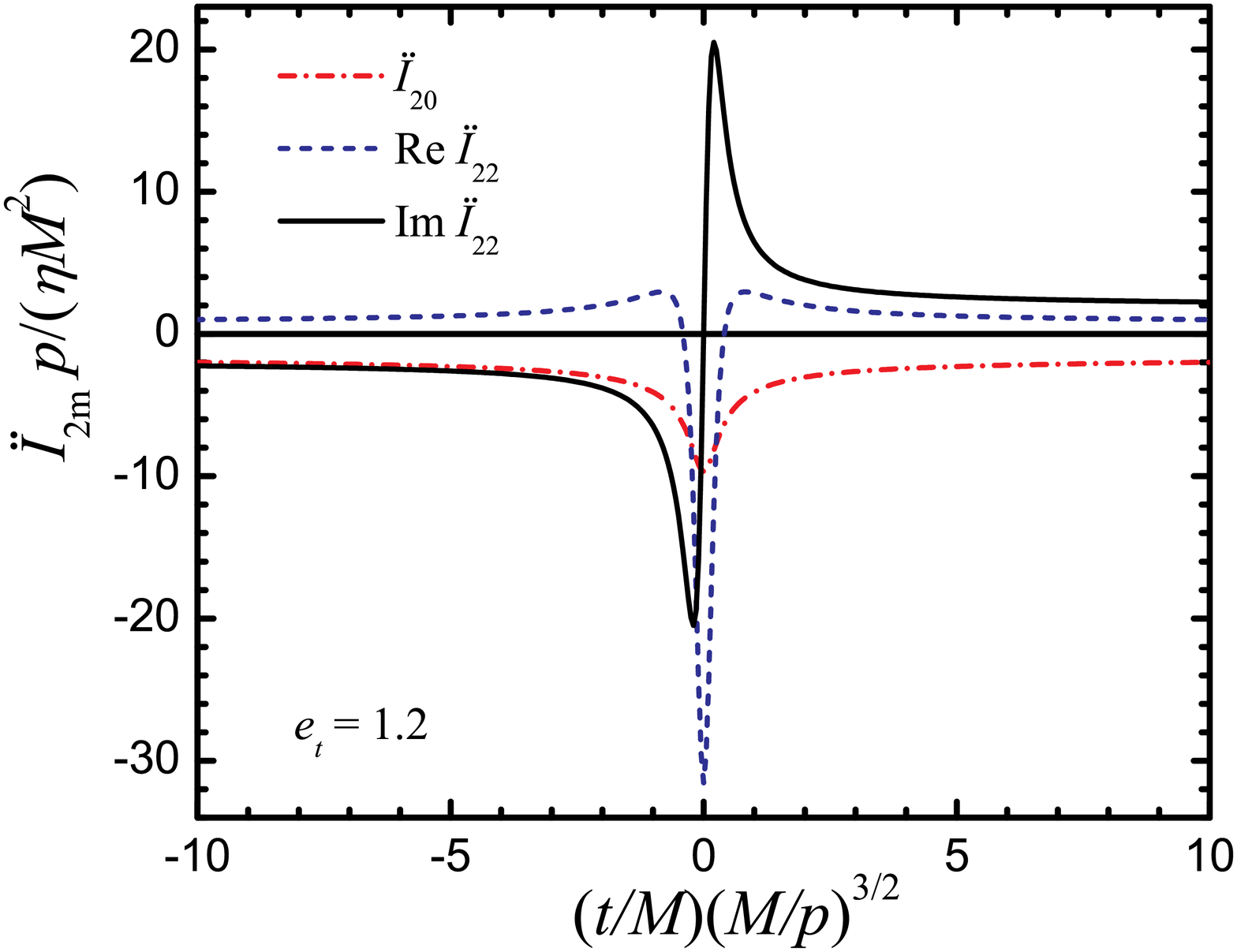} &
\includegraphics[angle=0, width=0.48\textwidth]{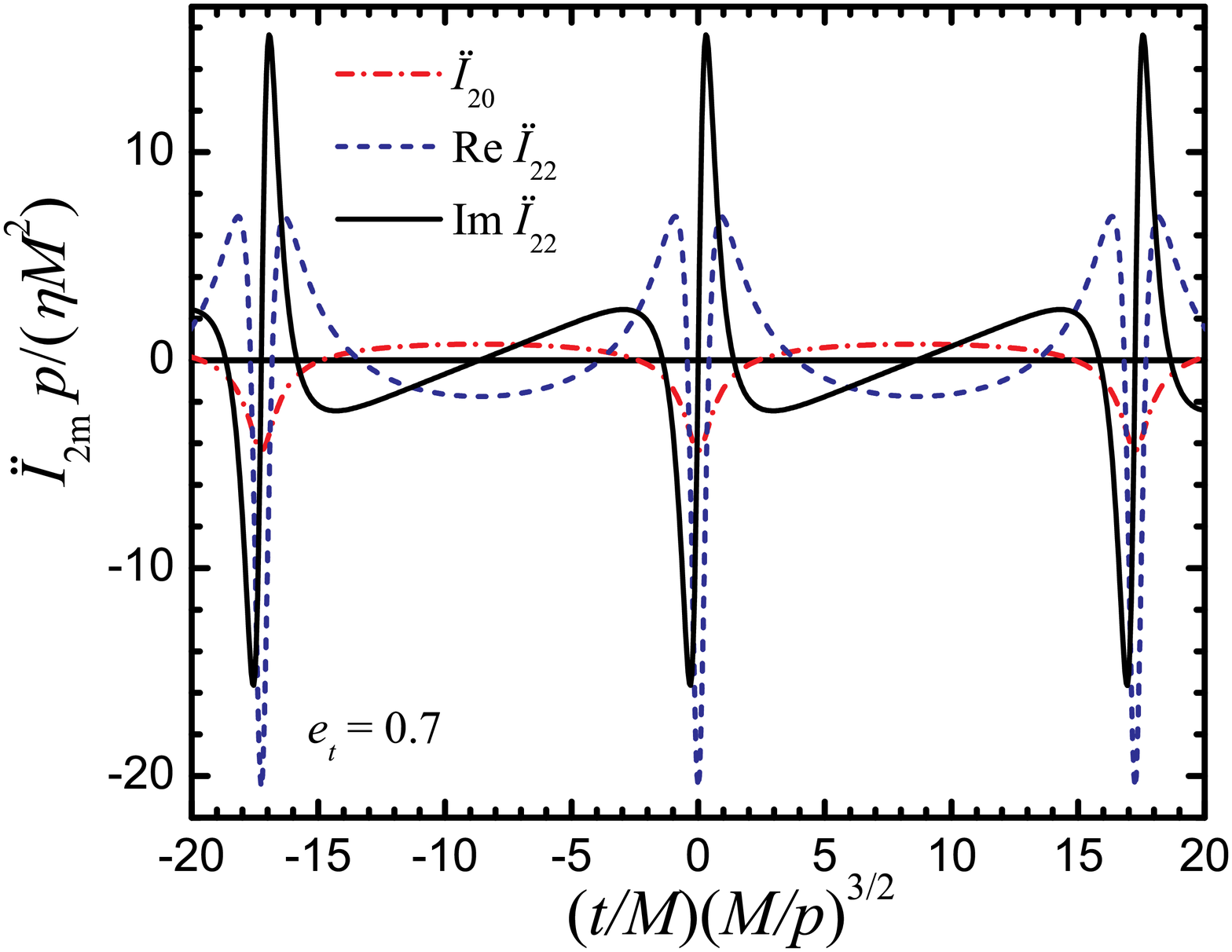}
\end{array}
$
\caption{\label{fig:Ilm-modes}(color online). Waveform modes for hyperbolic and eccentric orbits as a function of a normalized time coordinate $\hat{t} = (t/M) (M/p)^{3/2}$. Both plots show the $\ddot{I}_{lm}$ modes [Eqs.~\eqref{eq:d2I20dt2} and \eqref{eq:d2I22dt2}] that appear in the construction of the leading-order waveform polarizations. The eccentricity is indicated in each plot and completely determines the shape of the curves. I have assumed that the binary passes through the pericenter angle $\varpi=0$ at $\hat{t}=0$.  Note that the imaginary part of the $\ddot{I}_{22}$ mode shows a linear memory in the hyperbolic case (the size of this memory scales like $e_t/p \approx 1/b$ for large $e_t$ and impact parameter $b$; see Sec.~\ref{subsec:hyperbolicorbits}). Parabolic orbits ($e_t=0$) produce waveforms very similar to the hyperbolic ones, but all modes asymptote to the $\hat{t}$-axis at $t\rightarrow \pm \infty$ (no linear memory). The $\ddot{I}_{2-2}$ modes are similar to those plotted and are given by ${\rm Re} \ddot{I}_{2-2}(t)={\rm Re} \ddot{I}_{22}(t)$ and ${\rm Im} \ddot{I}_{2-2}(t)=-{\rm Im} \ddot{I}_{22}(-t)$.}
\end{figure*}

Applying Eqs.~\eqref{eq:keperlian} and \eqref{eq:kepderivatives}, Eqs.~\eqref{eq:dIlm} simplify to
\bs
\label{eq:d2I2mkep}
\be
\label{eq:d2I20dt2}
I^{(2)}_{20} = -8 \sqrt{\frac{\pi}{15}} \eta \frac{M^2}{p} e_t (e_t + \cos v) ,
\ee
\begin{multline}
\label{eq:d2I22dt2}
I^{(2)}_{2 \pm 2} = -4 \sqrt{\frac{2\pi}{5}} \eta \frac{M^2}{p} e^{\mp 2 i \varphi}
\\ \times
\left[ 1 - e_t^2 + (1+e_t \cos v) (1+ 2 e_t e^{\pm i v}) \right] ,
\end{multline}
\es
\bs
\be
\label{eq:d3I2mkep}
I^{(3)}_{20} = 8 \sqrt{\frac{\pi}{15}} \eta \left( \frac{M}{p} \right)^{5/2} e_t \sin v (1+e_t \cos v)^2 ,
\ee
\begin{multline}
I^{(3)}_{2 \pm 2} = -4 \sqrt{\frac{2 \pi}{5}} \eta \left( \frac{M}{p} \right)^{5/2} e^{\mp 2 i \varphi} (1+ e_t \cos v)^2
\\ \times
\left[ e_t \sin v \mp 4 i (1+ e_t \cos v) \right] .
\end{multline}
\es
The modes that appear in the waveform polarizations [Eqs.~\eqref{eq:d2I2mkep}] are plotted in Fig.~\ref{fig:Ilm-modes} for hyperbolic and elliptic orbits. To produce these plots, the differential equation for the true anomaly [\eqref{eq:dphidt}] was solved numerically.\footnote{These waveforms could also be produced by using a representation in terms of the \emph{eccentric anomaly} $u$. In this case, the problem of solving for the time involves finding a root of the so-called \emph{Kepler equation} (rather than solving a differential equation). However, in contrast to the true anomaly parameterization, separate sets of equations must be used to treat the elliptic, hyperbolic, and parabolic cases (see, e.g., Ch.~6 of \cite{danby}).} Note that the hyperbolic waveforms show a \emph{linear memory} in the $(2,\pm2)$ mode. This is discussed in more detail in Sec.~\ref{subsec:hyperbolicorbits}.

To compute the polarizations in terms of the true anomaly, one substitutes the following expressions into Eqs.~\eqref{eq:hpx}:
\bs
\begin{align}
&\dot{r} r \dot{\varphi} = \frac{M}{p} (1+ e_t \cos v) e_t \sin v,\\
&\dot{r}^2 + r^2 \dot{\varphi}^2 - \frac{M}{r} = \frac{M}{p} e_t (e_t + \cos v),\\
&\dot{r}^2 - r^2 \dot{\varphi}^2 - \frac{M}{r} = -\frac{M}{p} (2+ 3 e_t \cos v + e_t^2 \cos 2v).
\end{align}
\es
Likewise, we can use the above expressions for $r$, $\dot{r}$, and $\dot{\varphi}$ to write Eqs.~\eqref{eq:dhmem-newt} in terms of $e_t$, $p$, and $v$:
\begin{widetext}
\bs
\label{eq:dhmem-kep}
\begin{align}
\label{eq:dh20dt-kep}
h^{\rm (mem)(1)}_{20} &= \frac{32}{21 R} \sqrt{\frac{\pi}{30}} \eta^2 \bigg\langle \left( \frac{M}{p} \right)^5 (1+ e_t \cos v)^4
(24 + e_t^2 + 48 e_t \cos v + 23 e_t^2 \cos^2 v) \bigg\rangle ,\\
h^{\rm (mem)(1)}_{2 \pm 2} &= - \frac{16}{21 R} \sqrt{\frac{\pi}{5}} \eta^2 \bigg\langle  \left( \frac{M}{p} \right)^5 e^{\mp 2 i \varphi} e_t \sin v
(1+ e_t \cos v)^4 \left[ e_t \sin v \mp 4 i (1+e_t \cos v) \right] \bigg\rangle ,\\
h^{\rm (mem)(1)}_{40} &= \frac{2}{315 R} \sqrt{\frac{2 \pi}{5}} \eta^2 \bigg\langle  \left( \frac{M}{p} \right)^5 (1+ e_t \cos v)^4
(16 + 3 e_t^2 + 32 e_t \cos v + 13 e_t^2 \cos^2 v) \bigg\rangle ,\\
h^{\rm (mem)(1)}_{4 \pm 2} &= - \frac{4 \sqrt{\pi}}{315 R}  \eta^2 \bigg\langle  \left( \frac{M}{p} \right)^5 e^{\mp 2 i \varphi} e_t \sin v
(1+ e_t \cos v)^4 \left[ e_t \sin v \mp 4 i (1+e_t \cos v) \right] \bigg\rangle ,\\
h^{\rm (mem)(1)}_{4 \pm 4} &= - \frac{2}{45 R} \sqrt{\frac{\pi}{7}}  \eta^2 \bigg\langle  \left( \frac{M}{p} \right)^5 e^{\mp 4 i \varphi} (1+ e_t \cos v)^4
\left[ 16 - e_t^2 + 32 e_t \cos v + 17 e_t^2 \cos^2 v
 \pm 8 i e_t \sin v (1+e_t \cos v) \right] \bigg\rangle ,
\end{align}
\es
\end{widetext}
where the angle brackets again arise from the averaging inherent in the definition of the GW energy flux.
\subsection{\label{subsec:elliptical}Elliptical orbits}
For bound, eccentric orbits ($0 \leq e_t < 1$), the wavelength averaging in Eqs.~\eqref{eq:dhmem-kep} is accomplished by explicitly averaging over an orbital period $P_{\rm orb}$. For any function $F(t)$, this orbit-averaging is defined by
\begin{align}
\langle F(t) \rangle &= \frac{1}{P_{\rm orb}} \int_0^{P_{\rm orb}} dt \, F(t) \\
&= \frac{(1-e_t^2)^{3/2}}{2\pi} \int_0^{2\pi} dv \frac{F(v)}{(1+e_t \cos v)^2}, \nonumber
\end{align}
where
\be
\label{eq:Porb}
P_{\rm orb} = \frac{2\pi}{(1-e_t^2)^{3/2}} \sqrt{\frac{p^3}{M}}
\ee
follows from Eq.~\eqref{eq:time-eqn}.
Averaging Eqs.~\eqref{eq:dhmem-kep} then yields
\bs
\label{eq:dhlmdt-ecc}
\begin{align}
\label{eq:dh20dt-ecc}
h^{\rm (mem)(1)}_{20} \!\! &= \!\! \frac{256}{7 R} \sqrt{\! \frac{\pi}{30}} \eta^2 \! \left(\!\! \frac{M}{p}\!\! \right)^{\!\! 5} \!\! (1-e_t^2)^{3/2} \!\! \left( \!\! 1 \! + \! \frac{145}{48} e_t^2 \! + \! \frac{73}{192} e_t^4 \!\! \right) \!\! ,\\
h^{\rm (mem)(1)}_{2 \pm 2} \!\! &= \!\! \frac{52}{21 R} \sqrt{\! \frac{\pi}{5}} \eta^2 \! \left(\!\! \frac{M}{p}\!\! \right)^{\!\! 5} \!\! e_t^2 (1-e_t^2)^{3/2} \!\! \left( \!\! 1 \! + \! \frac{2}{13} e_t^2 \!\! \right) \! e^{\mp 2 i \varpi} \! ,\\
h^{\rm (mem)(1)}_{40} \!\! &= \!\! \frac{64}{315 R} \sqrt{\! \frac{\pi}{10}} \eta^2 \! \left(\!\! \frac{M}{p}\!\! \right)^{\!\! 5} \!\! (1-e_t^2)^{3/2} \!\! \left( \!\! 1 \! + \! \frac{99}{32} e_t^2  \! + \!\! \frac{51}{128} e_t^4 \!\! \right)  \!\! ,\\
h^{\rm (mem)(1)}_{4 \pm 2} \!\! &= \!\! \frac{13 \sqrt{\pi}}{315 R} \eta^2  \! \left(\!\! \frac{M}{p}\!\! \right)^{\!\! 5} \!\! e_t^2 (1-e_t^2)^{3/2} \!\! \left( \!\! 1 \! + \! \frac{2}{13} e_t^2 \!\! \right) e^{\mp 2 i \varpi} \! ,\\
h^{\rm (mem)(1)}_{4 \pm 4} \!\! &= \!\! -\frac{5}{72 R} \sqrt{\! \frac{\pi}{7}} \eta^2  \! \left(\!\! \frac{M}{p}\!\! \right)^{\!\! 5} \!\! e_t^4 (1-e_t^2)^{3/2}  e^{\mp 4 i \varpi} .
\end{align}
\es
This averaging has essentially removed the high frequency ($\sim 1/P_{\rm orb}$) structure from the memory waveform. Appendix \ref{app:averaging} explicitly shows that this averaging procedure has only a very small effect on the memory.

Next, we need to compute the time integrals of the above expressions. In the circular case treated in \cite{favata-pnmemory}, this was accomplished by changing variables to $x\equiv (M\omega)^{2/3}$. In the eccentric case, both the eccentricity $e_t$ and the semilatus rectum $p$ vary with time, but $p$ can be easily expressed in terms of $e_t$. We therefore change variables from time $t$ to eccentricity, and integrate from some early-time value of the eccentricity $e_{-}$ to its value at some later time $e_{+}=e_t(t)$:
\be
\label{eq:hlmetint}
h_{lm}^{\rm (mem)} = \int_{-\infty}^{T_R} h_{lm}^{{\rm (mem)}(1)} dt = \int_{e_{-}}^{e_{+}} \frac{h_{lm}^{{\rm (mem)}(1)}}{de_t/dt} d e_t.
\ee

In computing the above integral, we will need to make use of the following equations for the evolution of $p$ and $e_t$ [these are easily derived from the results of \cite{peters,DGI}, along with the Newtonian-order relations in Eqs.~\eqref{eq:Porb} and \eqref{eq:semilatusdefine}]:
\bs
\label{eq:dpdt-dedt}
\begin{align}
\label{eq:dpdt}
\frac{dp}{dt} &= -\frac{8}{5} \eta \left( \frac{M}{p} \right)^3 (1-e_t^2)^{3/2} (8+7e_t^2), \\
\label{eq:detdt}
\frac{de_t}{dt} &= -\frac{\eta}{15M} \left( \frac{M}{p} \right)^4 e_t (1-e_t^2)^{3/2} (304+121e_t^2).
\end{align}
\es
Dividing the first equation by the second yields
\be
\frac{dp}{de_t} = 24 \frac{p}{e_t} \frac{(8+7e_t^2)}{(304+121e_t^2)},
\ee
which can be solved to give
\be
\label{eq:p-et}
p(e_t) = \frac{p_0}{C_0} {e_t}^{12/19} ({304+121e_t^2})^{870/2299},
\ee
where
\be
C_0\equiv {e_0}^{12/19} ({304+121e_0^2})^{870/2299},
\ee
and $p_0$ is the value of $p$ at some arbitrary reference time when $e_t=e_0$. Since both $p$ and $e_t$ evolve with time, this relation allows us to eliminate the time-dependent $p(t)$ terms in Eq.~\eqref{eq:hlmetint} and instead express the integrand entirely in terms of the evolving eccentricity $e_t$ (which is our new integration variable) and the constants $p_0$ and $e_0$. The integrand in Eq.~\eqref{eq:hlmetint} for the relevant modes then becomes:
\bs
\label{eq:dhlmdet}
\begin{align}
\frac{dh_{20}^{\rm (mem)}}{de_t} &= -\frac{2}{7} \sqrt{\frac{10\pi}{3}} \frac{\eta M^2}{R p_0} \frac{C_0}{e_t^{31/19}} \frac{(192+580e_t^2 +73e_t^4)}{(304+121e_t^2)^{3169/2299}},\\
\frac{dh_{2\pm2}^{\rm (mem)}}{de_t} &= -\frac{4\sqrt{5\pi}}{7} \frac{\eta M^2}{R p_0} \frac{C_0 e_t^{7/19}(13+2e_t^2)}{(304+121e_t^2)^{3169/2299}} e^{\mp 2 i \varpi},\\
\frac{dh_{40}^{\rm (mem)}}{de_t} &= -\frac{1}{42} \sqrt{\frac{\pi}{10}} \frac{\eta M^2}{R p_0} \frac{C_0}{e_t^{31/19}} \frac{(128+396e_t^2 +51e_t^4)}{(304+121e_t^2)^{3169/2299}},\\
\frac{dh_{4\pm2}^{\rm (mem)}}{de_t} &= -\frac{\sqrt{\pi}}{21} \frac{\eta M^2}{R p_0} \frac{C_0 e_t^{7/19}(13+2e_t^2)}{(304+121e_t^2)^{3169/2299}} e^{\mp 2 i \varpi},\\
\frac{dh_{4\pm4}^{\rm (mem)}}{de_t} &= \frac{25}{24} \sqrt{\frac{\pi}{7}} \frac{\eta M^2}{R p_0} \frac{C_0 e_t^{45/19}}{(304+121e_t^2)^{3169/2299}} e^{\mp 4 i \varpi}.
\end{align}
\es
These expressions can be analytically integrated and the result expressed in terms of hypergeometric functions. Details of this are given in Appendix \ref{app:hypergeom}, where we show that all of the above modes can be expressed in the form
\be
\label{eq:hlm-hypergeom}
h_{lm}^{\rm (mem)} = A_{lm} C_0(e_0) e^{\mp i m \varpi} \left[ F_{lm}(e_t) - F_{lm}(e_{-}) \right],
\ee
where $A_{lm}$ are constants that can be read off of Eqs.~\eqref{eq:dhlmdet-app}, and $F_{lm}$ is a sum of hypergeometric functions given in Eq.~\eqref{eq:Flm}. Note that in computing the integral over $e_t$ we have chosen the integration constant such that the memory vanishes at an early-time eccentricity value of $e_{-}$. We have also ignored periastron precession (choosing $\varpi$ to be fixed), but the relaxation of this assumption will be discussed below.

In the limit of small $e_t$, we can easily evaluate the integrals of Eqs.~\eqref{eq:dhlmdet}:
\bs
\label{eq:hlm-smallet}
\begin{align}
\label{eq:h20-smallet}
h_{20}^{\rm (mem)} &= \frac{2}{7} \sqrt{\frac{10\pi}{3}} \frac{\eta M^2}{R p_0} \left[ \left( \frac{e_0}{e_t} \right)^{12/19} \!\!\!- \left( \frac{e_0}{e_{-}} \right)^{12/19} \right], \nonumber \\
&= \frac{2}{7} \sqrt{\frac{10\pi}{3}} \frac{\eta M^2}{R p} \left[ 1 - \left( \frac{e_t}{e_{-}} \right)^{12/19} \right],
\end{align}
\begin{align}
h_{2\pm2}^{\rm (mem)} &= -\frac{\sqrt{5\pi}}{56} \frac{\eta M^2}{R p_0} e^{\mp 2i\varpi} e_0^2 \! \left[ \! \left( \frac{e_t}{e_0} \right)^{26/19} \!\!\!\!- \left( \frac{e_{-}}{e_0} \right)^{26/19} \! \right]\!\!, \nonumber  \\
&= -\frac{\sqrt{5\pi}}{56} \frac{\eta M^2}{R p} e^{\mp 2i\varpi} e_t^2 \left[ 1 - \left( \frac{e_{-}}{e_t} \right)^{26/19} \right],
\end{align}
\begin{align}
h_{40}^{\rm (mem)} &= \frac{1}{63} \sqrt{\frac{\pi}{10}} \frac{\eta M^2}{R p_0} \left[ \left( \frac{e_0}{e_t} \right)^{12/19} \!\!\!- \left( \frac{e_0}{e_{-}} \right)^{12/19} \right], \nonumber  \\
&= \frac{1}{63} \sqrt{\frac{\pi}{10}} \frac{\eta M^2}{R p} \left[ 1 - \left( \frac{e_t}{e_{-}} \right)^{12/19} \right],
\end{align}
\begin{align}
h_{4\pm2}^{\rm (mem)} &= -\frac{\sqrt{\pi}}{672} \frac{\eta M^2}{R p_0} e^{\mp 2i\varpi} e_0^2 \! \left[ \! \left( \frac{e_t}{e_0} \right)^{26/19} \!\!\!\!- \left( \frac{e_{-}}{e_{0}} \right)^{26/19} \!  \right] \!\!, \nonumber  \\
&= -\frac{\sqrt{\pi}}{672} \frac{\eta M^2}{R p} e^{\mp 2i\varpi} e_t^2 \left[ 1 - \left( \frac{e_{-}}{e_t} \right)^{26/19} \right] ,
\end{align}
\begin{align}
h_{4\pm4}^{\rm (mem)} &= \frac{25\sqrt{7\pi}}{172\,032} \frac{\eta M^2}{R p_0} e^{\mp 4i\varpi} e_0^4 \! \left[ \! \left( \frac{e_t}{e_0} \right)^{64/19} \!\!\!\!- \left( \frac{e_{-}}{e_0} \right)^{64/19} \! \right] \!\!, \nonumber  \\
&= \frac{25\sqrt{7\pi}}{172\,032} \frac{\eta M^2}{R p} e^{\mp 4i\varpi} e_t^4 \left[ 1 - \left( \frac{e_{-}}{e_t} \right)^{64/19} \right] ,
\end{align}
\es
where we have used $C_0 \approx e_0^{12/19} 304^{870/2299}$, and in the second line of each equation we have re-expressed the result in terms of the time-dependent $p(t) \approx p_0 (e_t/e_0)^{12/19}$ using Eq.~\eqref{eq:p-et}.
\begin{figure*}[t]
$
\begin{array}{cc}
\includegraphics[angle=0, width=0.48\textwidth]{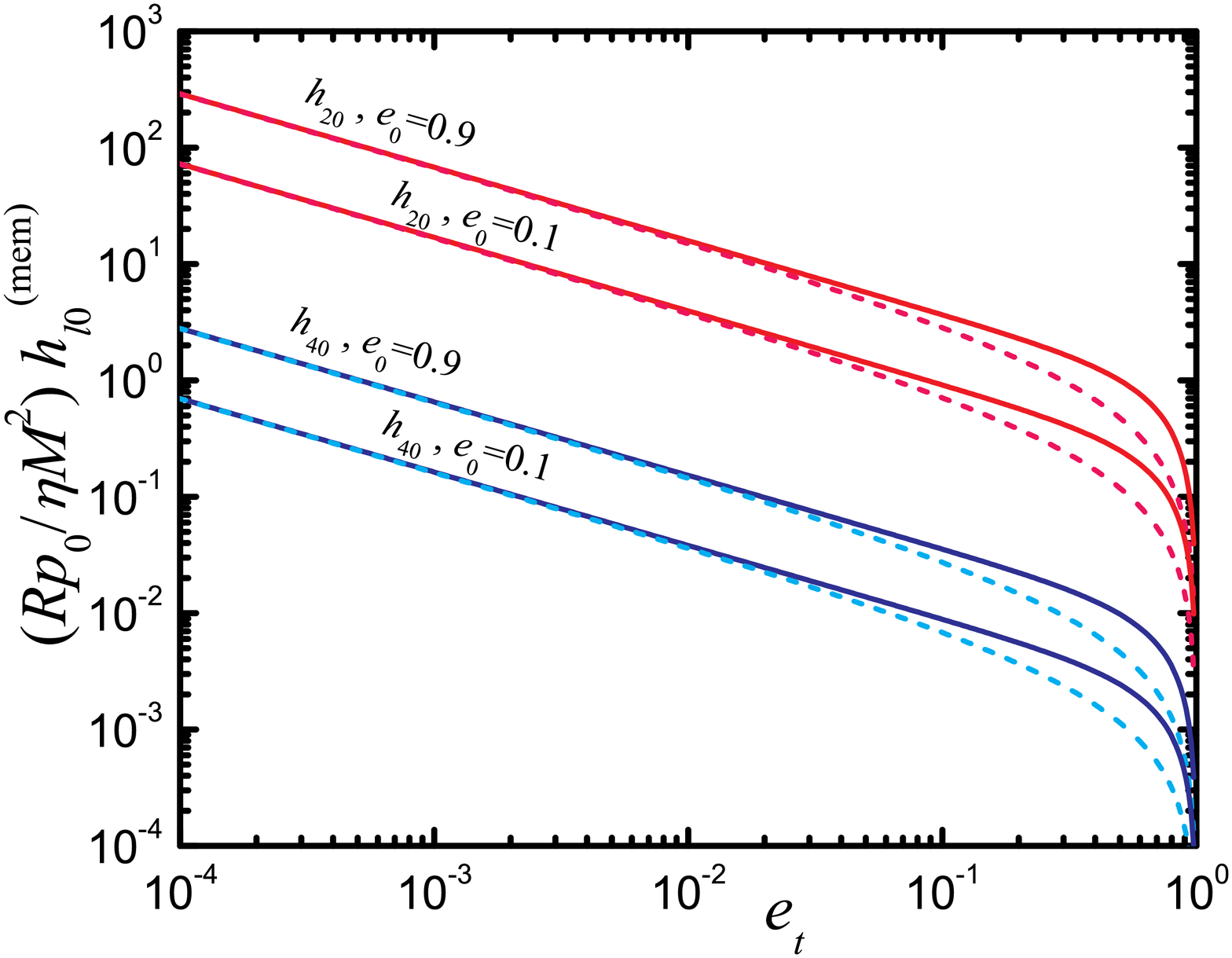} &
\includegraphics[angle=0, width=0.48\textwidth]{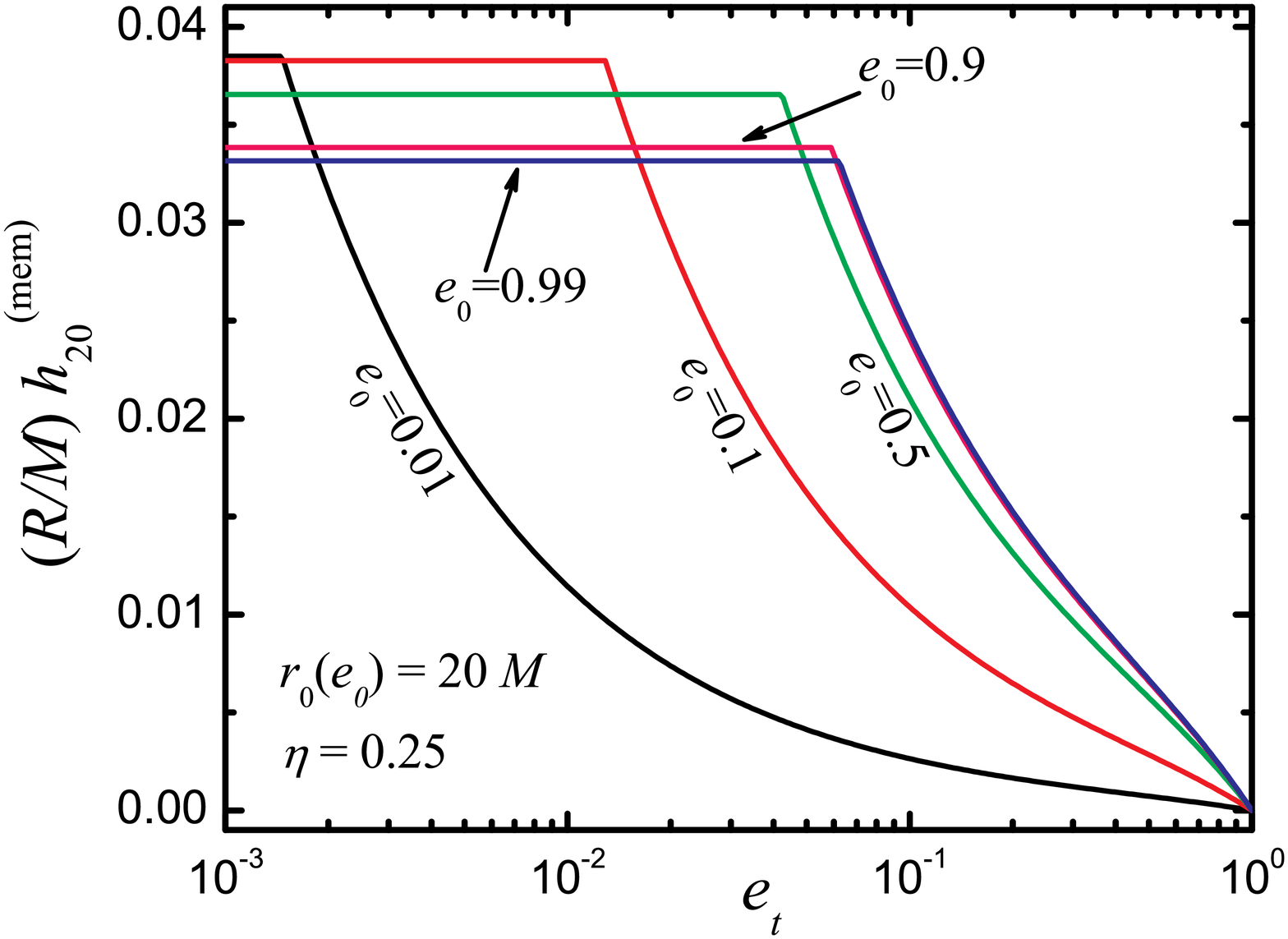}
\end{array}
$
\caption{\label{fig:hlmmodes}(color online). The left-hand plot shows the $h_{20}^{\rm (mem)}$ and $h_{40}^{\rm (mem)}$ memory modes as a function of eccentricity $e_t$ for two values of the reference eccentricity $e_0$ and for $e_{-}=1$. The dashed lines show the small-eccentricity approximation [Eqs.~\eqref{eq:hlm-smallet}]. In practice, these curves would not extend to arbitrarily small $e_t$ but would terminate when the last-stable-orbit is reached. The right-hand plot shows the evolution of the $h_{20}^{\rm (mem)}$ mode for various values of the indicated reference eccentricity $e_0$  (every curve passes through the point $e_t=e_0$ when $r_{\rm p}=20M$). The integration is terminated at the last-stable-orbit (at which point the curves flatten). In a real merger, the memory would continue to evolve past this point, eventually saturating at a different value.}
\end{figure*}

In the zero-eccentricity limit, $h_{2\pm2}^{\rm (mem)}$, $h_{4\pm2}^{\rm (mem)}$, and $h_{4\pm4}^{\rm (mem)}$ vanish and $h_{20}^{\rm (mem)}$ and $h_{40}^{\rm (mem)}$ reduce to the circular-orbit values found in Eqs.~(4.1)\mbox{--}(4.3) of \cite{favata-pnmemory}. In the elliptic case, the $m\neq 0$ modes do not contribute to the memory for several reasons: first, they are suppressed by factors of $e_t^m$ or $e_0^m$, and their numerical coefficients tend to be much smaller than the coefficient of $h_{20}^{\rm (mem)}$. More importantly, the factor of $e^{\mp i m \varpi}$ is not actually constant as we have assumed so far. Post-Newtonian corrections result in periastron precession, which causes $\varpi$ to vary with time [or with changing $e_t(t)$]. For example, the rate of periastron advance is
\be
\dot{\varpi} = \frac{2\pi k}{P_{\rm orb}} = \frac{3}{M} \left(\frac{M}{p} \right)^{5/2} (1-e_t^2)^{3/2}
\ee
for $k=3M/p$ at leading-PN-order. Using Eqs.~\eqref{eq:detdt} and \eqref{eq:p-et}, the pericenter angle can be obtained as a function of eccentricity by solving
\be
\frac{d\varpi}{de_t} = -\frac{45}{\eta} \left(\frac{p_0}{M} \right)^{3/2} \frac{C_0^{-3/2} e_t^{-1/19}}{(304+121e_t^2)^{994/2299}}.
\ee
For arbitrary (bound) eccentricities, this equation can be integrated using Eq.~\eqref{eq:hypergeomint}. For small eccentricity, this equation has the solution
\be
\varpi = \varpi_0 + \frac{5}{32\eta} \left( \frac{p_0}{M} \right)^{3/2} \left[ 1- \left( \frac{e_t}{e_0} \right)^{18/19} \right],
\ee
where $\varpi_0$ is the value of $\varpi$ when $e_t=e_0$. As the eccentricity varies from $e_0$ to $0$, the pericenter angle changes by $\Delta \varpi \sim (1/\eta)(p_0/M)^{3/2}$; this leads to oscillations in Eqs.~\eqref{eq:dhlmdet} that, upon integration, cause the $m\neq0$ terms to be further suppressed. Hence, as in the quasicircular case, only the $m=0$ terms contribute to a secularly increasing memory effect.

In the left-hand plot of Fig.~\ref{fig:hlmmodes}, we plot the $h_{20}^{\rm (mem)}$ and $h_{40}^{\rm (mem)}$ modes as a function of $e_t$. This is obtained from both the full analytic solution for the mode evolution [Eq.~\eqref{eq:hlm-hypergeom}, solid lines] and the low-eccentricity limit [Eqs.~\eqref{eq:hlm-smallet}, dashed lines], choosing $e_{-}=1$ in both cases. Note that the $h_{40}^{\rm (mem)}$ mode is much smaller than the $h_{20}^{\rm (mem)}$ mode.

It is also convenient to express the above results in terms of the pericenter distance $r_{\rm p}$ rather than $p$.
The time evolution of $r_{\rm p}$ is found from differentiating Eq.~\eqref{eq:rperi} and using Eqs.~\eqref{eq:dpdt-dedt}:
\be
\label{eq:drpdt}
\frac{dr_{\rm p}}{dt} = -\frac{\eta}{15} \! \left( \frac{M}{r_{\rm p}} \right)^3 \! \frac{(1-e_t)^{3/2}}{(1+e_t)^{7/2}} (192-112e_t +168 e_t^2 +47 e_t^3).
\ee
The evolution with eccentricity $r_{\rm p}(e_t)$ is easily found from Eq.~\eqref{eq:p-et},
\bs
\label{eq:rp-et}
\begin{align}
r_{\rm p} &= \frac{r_0}{C_0'} \frac{e_t^{12/19}}{(1+e_t)} (304+121 e_t^2)^{870/2299}\\
C_0' &\equiv \frac{e_0^{12/19}}{(1+e_0)} (304+121 e_0^2)^{870/2299}.
\end{align}
\es
This allows Eqs.~\eqref{eq:dhlmdt-ecc}, \eqref{eq:dhlmdet}, and \eqref{eq:hlm-smallet} to be expressed in terms of $r_{\rm p}$ or $r_0$.

The right-hand plot of Fig.~\ref{fig:hlmmodes} attempts to further illustrate the dependence of the $h_{20}^{\rm (mem)}$ memory mode for different eccentricities. In place of a time variable $t$, we can parameterize the temporal evolution in terms of the eccentricity $e_t$. This is because, at Newtonian order in the conservative dynamics, an inspiralling eccentric binary passes at some point in its evolution through every value of $e_t \in (0,1)$ with a one-to-one mapping between $t$ and $e_t$ (assuming we neglect the details of the binary's formation or its interactions with the external universe). To distinguish one eccentric binary from another, we need to specify the value of the eccentricity at some fiducial orbital separation. The different curves in the right-hand plot of Fig.~\ref{fig:hlmmodes} are parameterized by the value of the eccentricity $e_t=e_0$ when the binary passes through a pericenter distance of $r_{\rm p}=20M$. The curves are obtained from the analytic solution for the $(2,0)$ mode in Eq.~\eqref{eq:hlm-hypergeom}, choosing $e_{-}=1$. The $(2,0)$ mode is allowed to grow until the last-stable-orbit (LSO) is reached, corresponding to the condition $p\equiv r_{\rm p} (1+e_t) =6+2e_t$ \cite{cutler-kennefick-poisson} [the LSO value of eccentricity $e_{\rm LSO}$ is determined by combining this condition with Eq.~\eqref{eq:rp-et}]. This plot shows that binaries with large $e_0$ reach the LSO while they are still mildly eccentric and at slightly smaller values of the memory. (However, note that GWs radiated during the merger and ringdown will cause the memory to grow past its LSO value.)

The polarizations for the nonlinear-memory waves for bound, eccentric orbits can be simply computed by summing the $m=0$ modes in Eq.~\eqref{eq:hdecompose}:
\bs
\begin{align}
h_{+}^{\rm (mem)} &= \frac{1}{8} \sqrt{\frac{30}{\pi}} {\rm s}^2_{\Theta} \left[ h_{20}^{\rm (mem)} + \frac{\sqrt{3}}{2} h_{40}^{\rm (mem)} (7 {\rm c}^2_{\Theta} -1) \right], \\
h_{\times}^{\rm (mem)} &= 0.
\end{align}
\es
For arbitrary eccentricities, Eq.~\eqref{eq:hlm-hypergeom} for the $l=2$ and $l=4$ modes must be substituted into the above equation. In the small-eccentricity case, Eqs.~\eqref{eq:hlm-smallet} yield
\begin{align}
\label{eq:hpmemsmallet}
h_{+}^{\rm (mem)} &= \frac{\eta}{48} \frac{M^2}{R p_0} {\rm s}^2_{\Theta} (17 + {\rm c}^2_{\Theta}) \! \left[ \! \left( \frac{e_0}{e_t} \right)^{12/19} \!\!\!\!- \left( \frac{e_0}{e_{-}} \right)^{12/19} \right]\!\!, \nonumber \\
&= \frac{\eta}{48} \frac{M^2}{R p} {\rm s}^2_{\Theta} (17 + {\rm c}^2_{\Theta}) \left[ 1 - \left( \frac{e_t}{e_{-}} \right)^{12/19} \right],
\end{align}
Note that in the circular limit this agrees with Eqs.~(4.4)\mbox{--}(4.6a) of \cite{favata-pnmemory} or Eq.~\eqref{eq:hplus-intro} above.
\subsection{\label{subsec:hyperbolicorbits}Hyperbolic and parabolic orbits}
To treat the case of hyperbolic and parabolic orbits, we ignore the possibility of periastron advance and we fix the periastron direction to lie along the $+x$ axis. In this case, the reduced mass particle swings around the origin in a counterclockwise sense, entering at very early times along the asymptote at $\varphi=v_{-}\equiv -\arccos(-1/e_t)$, and exiting at very late times along the asymptote at $\varphi=v_{+}\equiv \arccos(-1/e_t)$ (see Fig.~\ref{fig:diagrams}). The corresponding scattering angle $\Theta_s$ is given by
\be
\Theta_s= 2 \arccos(-1/e_t) -\pi.
\ee
For hyperbolic orbits, it is also useful to define two additional parameters that can be used in place of $p$ or $e_t$. The asymptotic velocity is
\be
V_{\infty}^2 = 2 \frac{E}{\mu} = \frac{M}{p} (e_t^2-1).
\ee
The impact parameter $b$, defined to be the perpendicular distance from the center of mass $M$ to the ingoing or outgoing asymptote of the hyperbola, is found to be
\be
b = \frac{p}{\sqrt{e_t^2-1}}.
\ee
The equations below can alternatively be expressed in terms of $V_{\infty}$ or $b$ using the above relations.

It is important to note that the waveforms from hyperbolic orbits already contain a linear memory \cite{turner-unbound}. For example, consider Eqs.~\eqref{eq:d2I20dt2} and \eqref{eq:d2I22dt2} for $e_t>1$ and $\varphi=v$ varying between $v_{-}$ and $v_{+}$. This difference in the orbital phase angle does not affect the $I_{20}^{(2)}$ mode (since $\cos v$ is even), but it does affect the imaginary part of $I_{2\pm2}^{(2)}$ (which is odd), leading to a memory in that mode (see Fig.~\ref{fig:Ilm-modes}). More explicitly, the linear memory jump between late and early times for a hyperbolic orbit is found from the difference $I_{2m}^{(2)}(v_{+}) - I_{2m}^{(2)}(v_{-})$, yielding [see also Eqs.~(10) of \cite{favata-amaldiconfproc-memory-CQG2010}]
\bs
\label{eq:Deltah-linear}
\begin{align}
\Delta h_{20}^{\rm (lin.~mem)} &= 0, \\
\Delta h_{2\pm2}^{\rm (lin.~mem)} &= \pm i 16 \sqrt{\frac{\pi}{5}} \eta \frac{M}{R}\frac{M}{p} \frac{(e_t^2-1)^{3/2}}{e_t^2}.
\end{align}
\es
The corresponding memory jump in the polarizations is
\bs
\label{eq:Deltahpx-linear}
\begin{align}
\Delta h_{+}^{\rm (lin.~mem)} &= -4 \eta \frac{M}{R} \frac{M}{p} \frac{(e_t^2-1)^{3/2}}{e_t^2} (1+{\rm c}^2_{\Theta}) \sin2\Phi,\\
\Delta h_{\times}^{\rm (lin.~mem)} &= -8 \eta \frac{M}{R} \frac{M}{p} \frac{(e_t^2-1)^{3/2}}{e_t^2} {\rm c}_{\Theta} \cos2\Phi.
\end{align}
\es
Note that for large $e_t$,\footnote{In this case, we have $V_{\infty}^2 \approx e_t^2 (M/p)$ and $b\approx p/e_t$, or alternatively, $e_t\approx V_{\infty}^2 (b/M)$ and $p\approx V_{\infty}^2 (b^2/M)$. The $e_t\gg1$ limit then corresponds to the bremsstrahlung (small-angle scattering) limit, $V_{\infty}^2 \gg M/b$. Note that the scattering angle for $e_t\gg 1$ is $\Theta_s \approx 2/e_t \approx (2/V_{\infty}^2)(M/b) \ll 1$.}
\be
\frac{M}{p} \frac{(e_t^2-1)^{3/2}}{e_t^2} \approx \frac{M}{b},
\ee
and Eqs.~\eqref{eq:Deltahpx-linear} agree with Eqs.~(15) of \cite{wiseman-will-memory}.
Note also that for parabolic orbits ($e_t=1$), the linear memory vanishes. This is because the asymptotic incoming and outgoing directions of the orbit are now the same ($v_{-}=-\pi$, $v_{+}=+\pi$).

\begin{figure}[t]
\includegraphics[angle=0, width=0.48\textwidth]{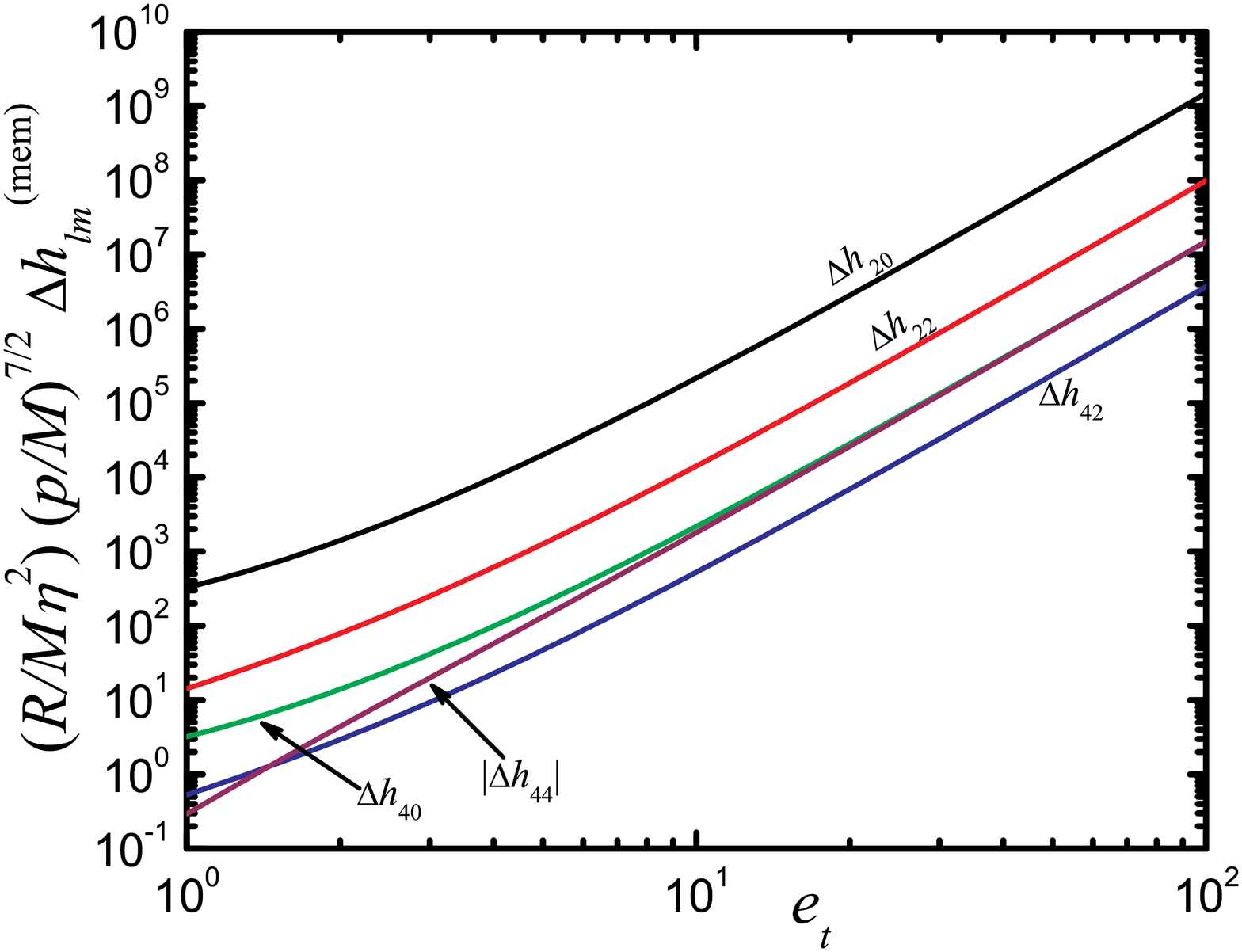}
\caption{\label{fig:hlmmodes-hyperbolic}(color online). The $\Delta h_{lm}^{\rm (mem)}$ modes for hyperbolic orbits from Eqs.~\eqref{eq:Deltahlm-hyperbolic}. For $e_t=1$, these reduce to Eqs.~\eqref{eq:Deltahlm-parabolic}, while for $e_t\gg1$, they asymptote to Eqs.~\eqref{eq:Deltahlm-largeet}. }
\end{figure}

To compute the nonlinear memory we proceed from Eqs.~\eqref{eq:dhmem-kep}. Since unbound orbits are no longer periodic, there is no need to average over an orbital period. Instead, we directly perform the time integrals over Eq.~\eqref{eq:dhmem-kep}, changing variables to the true anomaly using Eq.~\eqref{eq:dphidt}:
\be
h_{lm}^{\rm (mem)} = \int_{v_{-}}^{v(T_R)} \frac{h_{lm}^{\rm (mem)(1)}}{\dot{v}} dv.
\ee
The integrand is a sum over powers of sines and cosines or their products, and is easily evaluated for any limit of integration. For simplicity (and because periastron passage happens relatively quickly for hyperbolic orbits), we will focus on computing only the overall memory jump $\Delta h_{lm}^{\rm (mem)}$, rather than the evolution of the memory with time:
\be
\Delta h_{lm}^{\rm (mem)} = \int_{v_{-}}^{v_{+}} \frac{h_{lm}^{\rm (mem)(1)}}{\dot{v}} dv.
\ee
The resulting memory modes for any $e_t\geq 1$ are
\begin{widetext}
\bs
\label{eq:Deltahlm-hyperbolic}
\begin{align}
\Delta h_{20}^{\rm (mem)} &= \frac{8}{63} \sqrt{\frac{\pi}{30}} \eta^2 \frac{M}{R} \left(\frac{M}{p}\right)^{7/2} \left[ 3(73e_t^4+580e_t^2+192)(\pi-\arccos{e_t}^{-1}) + (1333e_t^2+1202)\sqrt{e_t^2-1} \right],\\
\Delta h_{2\pm2}^{\rm (mem)} &= \frac{8}{63} \sqrt{\frac{\pi}{5}} \eta^2 \frac{M}{R} \left(\frac{M}{p}\right)^{7/2} \left[ 3e_t^2(2e_t^2+13)(\pi-\arccos{e_t}^{-1}) + (34e_t^2+13-2/e_t^2)\sqrt{e_t^2-1} \right],\\
\Delta h_{40}^{\rm (mem)} &= \frac{1}{945} \sqrt{\frac{\pi}{10}} \eta^2 \frac{M}{R} \left(\frac{M}{p}\right)^{7/2} \left[ 3(51e_t^4+396e_t^2+128)(\pi-\arccos{e_t}^{-1}) + (919e_t^2+806)\sqrt{e_t^2-1} \right],\\
\Delta h_{4\pm2}^{\rm (mem)} &= \frac{2\sqrt{\pi}}{945} \eta^2 \frac{M}{R} \left(\frac{M}{p}\right)^{7/2} \left[ 3e_t^2(2e_t^2+13)(\pi-\arccos{e_t}^{-1}) + (34e_t^2+13-2/e_t^2)\sqrt{e_t^2-1} \right],\\
\Delta h_{4\pm4}^{\rm (mem)} &= -\frac{1}{2700} \sqrt{\frac{\pi}{7}} \eta^2 \frac{M}{R} \left(\frac{M}{p}\right)^{7/2} \left[ 375e_t^4 (\pi-\arccos{e_t}^{-1}) + (1001e_t^2-1178+728/e_t^2 -176/e_t^4)\sqrt{e_t^2-1} \right].
\end{align}
\es
Note that there is a memory contribution of order $\eta^2 (M/p)^{7/2}$ in each mode (a relative 2.5PN correction). Note also that each mode is real-valued. These modes are plotted in Fig.~\ref{fig:hlmmodes-hyperbolic}. The resulting polarizations are
\bs
\label{eq:hyperbolic-polarizations}
\begin{multline}
\Delta h_{+}^{\rm (mem)} = \frac{\eta^2}{960} \frac{M}{R} \left(\frac{M}{p}\right)^{7/2} \Bigg\{ ({\rm c}^4_{\Theta}-1) \left[ 50 e_t^4 (\pi-\arccos e_t^{-1}) + \frac{\sqrt{e_t^2-1}}{15} \left(2002 e_t^2-2356+\frac{1456}{e_t^2}-\frac{352}{e_t^4} \right) \right] \cos4\Phi
\\
+\frac{32}{3} (3+2 {\rm c}^2_{\Theta} + {\rm c}^4_{\Theta}) \left[ 3 e_t^2 (2e_t^2+13)(\pi-\arccos e_t^{-1}) + \sqrt{e_t^2-1} \left(34e_t^2+13-\frac{2}{e_t^2}\right) \right] \cos2\Phi
\\
+4(\pi-\arccos e_t^{-1}) \left[ (827e_t^4+6572e_t^2+2176) -(776e_t^4+6176e_t^2+2048) {\rm c}^2_{\Theta} -(51e_t^4+396e_t^2+128) {\rm c}^4_{\Theta} \right]
\\
+\frac{4}{3} \sqrt{e_t^2-1} \left[ 15103e_t^2+13622-(14184e_t^2+12816) {\rm c}^2_{\Theta} -(919e_t^2+806) {\rm c}^4_{\Theta} \right] \Bigg\},
\end{multline}
\begin{multline}
\Delta h_{\times}^{\rm (mem)} = \frac{\eta^2}{90} \frac{M}{R} \left(\frac{M}{p}\right)^{7/2} {\rm c}_{\Theta} \left\{ ({\rm s}^2_{\Theta} -6) \left[ e_t^2 (6e_t^2+39) (\pi - \arccos e_t^{-1}) + \sqrt{e_t^2-1} \left( 34e_t^2 +13 -\frac{2}{e_t^2} \right) \right] \sin2\Phi
\right.\\ \left.
+\frac{{\rm s}^2_{\Theta}}{40} \left[ 375e_t^4 (\pi-\arccos e_t^{-1}) + \sqrt{e_t^2-1} \left( 1001e_t^2-1178 + \frac{728}{e_t^2} - \frac{176}{e_t^4} \right) \right] \sin4\Phi \right\}.
\end{multline}
\es
\end{widetext}

For parabolic orbits ($e_t=1$), Eqs.~\eqref{eq:Deltahlm-hyperbolic} simplify to:
\bs
\label{eq:Deltahlm-parabolic}
\begin{align}
\Delta h_{20, e_t=1}^{\rm (mem)} &= \frac{676\pi\sqrt{30 \pi}}{63} \eta^2 \frac{M}{R} \left(\frac{M}{p}\right)^{7/2}, \\
\Delta h_{2\pm2, e_t=1}^{\rm (mem)} &= \frac{3\sqrt{6}}{169} \Delta h_{20, e_t=1}^{\rm (mem)},\\
\Delta h_{40, e_t=1}^{\rm (mem)} &= \frac{23\sqrt{3}}{4056} \Delta h_{20, e_t=1}^{\rm (mem)},\\
\Delta h_{4\pm2, e_t=1}^{\rm (mem)} &= \frac{\sqrt{30}}{3380} \Delta h_{20, e_t=1}^{\rm (mem)},\\
\Delta h_{4\pm4, e_t=1}^{\rm (mem)} &= -\frac{\sqrt{210}}{16224} \Delta h_{20, e_t=1}^{\rm (mem)},
\end{align}
\es
and the corresponding polarizations are
\bs
\label{eq:Deltahpx-parabolic}
\begin{multline}
\Delta h_{+, e_t=1}^{\rm (mem)}= \frac{\pi \eta^2}{2} \frac{M}{R} \left(\frac{M}{p}\right)^{7/2}
\bigg\{ (3+2 {\rm c}^2_{\Theta} + {\rm c}^4_{\Theta})\cos2\Phi
 \\
+ \frac{5}{48} {\rm s}^2_{\Theta} \left[ 766 + 46 {\rm c}^2_{\Theta} - (1+{\rm c}^2_{\Theta}) \cos4\Phi \right] \bigg\},
\end{multline}
\be
\Delta h_{\times, e_t=1}^{\rm (mem)} \! = \! \frac{\pi \eta^2}{2} \frac{M}{R} \! \left( \!\! \frac{M}{p} \!\! \right)^{\!\! 7/2} \!\!\! {\rm c}_{\Theta}
\!\! \left[\! \frac{5}{24} {\rm s}^2_{\Theta} \sin4\Phi \! - \! (5 \!+ \! {\rm c}^2_{\Theta})\sin2\Phi \! \right] \!\! .
\ee
\es
Unlike in the linear-memory case, the nonlinear memory for parabolic orbits is nonzero. Even though parabolic orbits are marginally bound, their radiated GWs are unbound and hence contribute to the nonlinear memory.

We can also examine the $e_t\gg 1$ limit. In this case, it is easy to extract the large-$e_t$ behavior from Eqs.~\eqref{eq:Deltahlm-hyperbolic}:
\bs
\label{eq:Deltahlm-largeet}
\begin{align}
\Delta h_{20, e_t\gg1}^{\rm (mem)} &= \frac{292\pi}{21}\sqrt{\frac{\pi}{30}} \eta^2 \frac{M}{R} \left(\frac{M}{p}\right)^{7/2} e_t^4, \\
\Delta h_{2\pm2, e_t\gg1}^{\rm (mem)} &= \frac{2\sqrt{6}}{73} \Delta h_{20, e_t\gg1}^{\rm (mem)}, \\
\Delta h_{40, e_t\gg1}^{\rm (mem)} &= \frac{17\sqrt{3}}{2920} \Delta h_{20, e_t\gg1}^{\rm (mem)}, \\
\Delta h_{4\pm2, e_t\gg1}^{\rm (mem)} &= \frac{\sqrt{30}}{2190} \Delta h_{20, e_t\gg1}^{\rm (mem)}, \\
\Delta h_{4\pm4, e_t\gg1}^{\rm (mem)} &= -\frac{5\sqrt{210}}{7008} \Delta h_{20, e_t\gg1}^{\rm (mem)}.
\end{align}
\es
Using
\be
\left(\frac{M}{p}\right)^{7/2} e_t^4 \approx \left(\frac{M}{b}\right)^3 V_{\infty},
\ee
the polarizations are given by
\bs
\label{eq:Deltahpx-hyperbolic}
\begin{multline}
\Delta h_{+, e_t\gg1}^{\rm (mem)} = \frac{\pi}{960} \eta^2 \frac{M}{R} \left(\frac{M}{b}\right)^3 V_{\infty}
\\ \times
 \left[ 192\cos2\Phi + {\rm s}^2_{\Theta} (1756-128\cos2\Phi-50\cos4\Phi)
\right. \\ \left.
  - {\rm s}^4_{\Theta} (102-32\cos2\Phi-25\cos4\Phi) \right],
\end{multline}
\begin{multline}
\Delta h_{\times, e_t\gg1}^{\rm (mem)} = -\frac{\pi}{480} \eta^2 \frac{M}{R} \left(\frac{M}{b}\right)^3 V_{\infty} {\rm c}_{\Theta}
\\ \times
\left[ 96\sin2\Phi - {\rm s}^2_{\Theta} (16\sin2\Phi+25\sin4\Phi) \right].
\end{multline}
\es
This agrees exactly with Eq.~(16) of \cite{wiseman-will-memory}, providing further confirmation of the correctness of the above results.

Note also the different scalings between the linear [Eq.~\eqref{eq:Deltah-linear}] and nonlinear memories in the $e_t\gg1$ limit:
\begin{align}
\Delta h^{\rm lin.\,mem}_{+,\times, e_t\gg1} &\propto \eta \frac{M}{R} \frac{M}{b},\\
\Delta h^{\rm nonlin.\,mem}_{+,\times, e_t\gg1} &\propto \eta^2 \frac{M}{R} \left(\frac{M}{b}\right)^{3} V_{\infty}
\end{align}
This indicates that the nonlinear memory for high-velocity gravitational scattering is typically much smaller than the linear memory (see also Sec.~\ref{sec:estimates}).
This is in contrast to the case of bound eccentric (and circular) orbits, where the linear memory vanishes (but see Sec.~V B of \cite{favata-pnmemory} for a caveat) while the nonlinear memory is $\propto \eta$. These scaling differences in the nonlinear memory arise from differences in the integration time over which the nonlinear memory builds up [see the discussion following Eq.~\eqref{eq:hmemintegral} above].
\subsection{\label{subsec:radial}Radial orbits}
Next we consider radial orbits corresponding to the head-on collision or separation of two masses. In this case, the equations of motion and conserved energy yield
\bs
\be
\dot{\varphi}=\ddot{\varphi}=0, \;\;\;\; \ddot{r} = -\frac{M}{r^2} ,
\ee
\be
\label{eq:E-rdot-eqn}
\tilde{E} \equiv \frac{E}{\mu} = \frac{\dot{r}^2}{2} - \frac{M}{r}.
\ee
\es
The multipole modes in Eqs.~\eqref{eq:dIlm} easily simplify in the radial case (where we can choose $\varphi=\text{const}=0$), and the leading-order waveform polarizations become
\bs
\be
h_+^{\rm N} = \frac{\eta M}{R} \left\{ \left( \dot{r}^2 - \frac{M}{r} \right) \left[ (1+c_{\Theta}^2) \cos2\Phi - s^2_{\Theta} \right] \right\},
\ee
\be
h_{\times}^{\rm N} = -\frac{2\eta M}{R} \left( \dot{r}^2 - \frac{M}{r} \right) c_{\Theta} \sin2\Phi .
\ee
\es
If the relative radial velocity approaches $v_{\infty}$ at infinite separation, $\dot{r}^2 - M/r \rightarrow v_{\infty}^2+M/r$. Radial waveforms can therefore show a \emph{linear} memory effect that depends on $v_{\infty}$ and the initial and final values of $M/r$.

To compute the nonlinear memory, we simplify Eqs.~\eqref{eq:dhmem-newt} (again choosing $\varphi=0)$. We easily see that all of the leading-order memory modes have the form
\be
h_{lm}^{(\rm mem)(1)} = C_{lm} \frac{\eta^2}{R} \left(\frac{M}{r}\right)^4 \dot{r}^2,
\ee
where the constants $C_{lm}$ can be read off of Eqs.~\eqref{eq:dhmem-newt}.
Converting the time-integral of $h_{lm}^{(\rm mem)(1)}$ to a radial integral and using Eq.~\eqref{eq:E-rdot-eqn},
\be
\dot{r} = \pm \sqrt{2} \sqrt{\tilde{E} + M/r},
\ee
(where the upper sign here and below indicates increasing radial separation and lower sign indicates radial infall), the $h_{lm}^{(\rm mem)}$ modes can be expressed as
\be
h_{lm}^{(\rm mem)} = \pm \sqrt{2} C_{lm} \frac{\eta^2}{R} \int_{r_{-}}^{r_{+}} \left( \frac{M}{r} \right)^4 \sqrt{\tilde{E} + \frac{M}{r}} dr,
\ee
where $r_{\pm}$ refers to the value of $r$ at late or early times. Evaluating the integral yields
\begin{multline}
\label{eq:hlmradial1}
h_{lm}^{(\rm mem)} = \mp \frac{2\sqrt{2}}{105} C_{lm} \frac{\eta^2 M}{R} \left\{ \left( \tilde{E} + \frac{M}{r} \right)^{3/2}
\right. \\ \left. \times \left[ 8 \tilde{E}^2 - 12 \tilde{E} \frac{M}{r} + 15 \left(\frac{M}{r}\right)^2 \right] \right\} \Bigg|_{r_{-}}^{r_+}.
\end{multline}

For the case of radial infall from rest at infinity, the above simplifies to
\be
\label{eq:hlmradial2}
h_{lm}^{(\rm mem)} = \frac{2\sqrt{2}}{7} C_{lm} \frac{\eta^2 M}{R} \left[\frac{M}{r(t)}\right]^{7/2},
\ee
where $r_+ \rightarrow r(t)$.
We can also consider the case of radial binary disruption (e.g., this could also model a star that radially ejects a piece of material). If the initial separation is $r_{-}$ and the asymptotic late-time relative velocity of the two components approaches $v_{\infty}$ (so that $\tilde{E}=v_{\infty}^2/2$), the resulting nonlinear memory shift is
\begin{multline}
\label{eq:hlmradial3}
\Delta h_{lm}^{(\rm mem)} = - \frac{2}{105} C_{lm} \frac{\eta^2 M}{R} \left[ v_{\infty}^7 - \sqrt{2} \left( \frac{v_{\infty}^2}{2} + \frac{M}{r_{-}} \right)^{3/2}
\right. \\ \left.
\times \left( 2 v_{\infty}^4 - 6 v_{\infty}^2 \frac{M}{r_{-}} + 15 \frac{M^2}{r_{-}^2} \right) \right].
\end{multline}
Note that in both of the above cases the nonlinear memory is a relative 2.5PN correction to the Newtonian waveform.
In all cases, the waveform polarizations for radial orbits are given explicitly by
\bs
\begin{multline}
h_{+}^{\rm (mem)} = \hat{h}^{(\rm mem)} \left[ \frac{s^2_{\Theta}}{420} (79+ 7 c^2_{\Theta})
\right. \\ \left.
 - \frac{1}{15} (3+ 2c^2_{\Theta} + c^4_{\Theta}) \cos2\Phi + \frac{1}{60} (1-c^4_{\Theta}) \cos4\Phi \right],
\end{multline}
\be
h_{\times}^{\rm (mem)} = \hat{h}^{(\rm mem)} \left[ \frac{c_{\Theta}}{15} (5+ c^2_{\Theta}) \sin2\Phi -\frac{1}{30} s^2_{\Theta} c_{\Theta} \sin4\Phi \right],
\ee
\es
where $\hat{h}^{(\rm mem)}$ is given by Eqs.~\eqref{eq:hlmradial1}, \eqref{eq:hlmradial2}, or \eqref{eq:hlmradial3} with $\hat{h}^{(\rm mem)} \equiv h_{lm}^{\rm (mem)} /C_{lm}$.
\section{\label{sec:earlymemory}Sensitivity of the memory to the early-time history of a binary}
\begin{figure}[t]
\includegraphics[angle=0, width=0.48\textwidth]{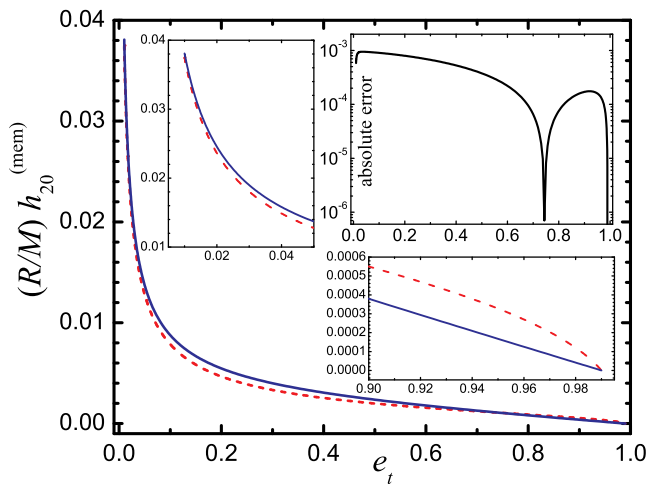}
\caption{\label{fig:h20-circ-ecc}(color online). Evolution of the $(2,0)$ memory mode in the quasicircular (red/dashed) and elliptical (blue/solid) cases for an equal-mass binary. The two modes are evolved as described in the main text. Time for \emph{both} curves is parameterized by the eccentricity of the elliptic case. The elliptic mode is set to zero at an eccentricity of $e_{-}=0.99$ and evolves until it reaches a value of $e_0=0.01$ at $r_{\rm p}=6M$. The quasicircular mode also evolves up to a separation of $6M$ and is set to zero at the time when the eccentricity of the elliptic mode is $0.99$. Two of the insets zoom in on the early- and late-time stages of the evolution (at high and low eccentricities, respectively). The other inset shows the absolute value of the difference between the two curves, $|h_{20}^{\rm (mem, circ)}-h_{20}^{\rm (mem, ellip)}|$.}
\end{figure}

In this section, we wish to evaluate the degree to which the nonlinear memory from a quasicircular inspiralling binary is sensitive to its deviations from circularity. These deviations arise from the binary's initial eccentricity, and are damped by radiation reaction (in absence of external perturbing forces). To perform this evaluation, we compare two models for the evolution of the $h_{20}^{\rm (mem)}$ mode. (For simplicity and because they tend to be much smaller, I will neglect the other memory modes.) In the first model, we consider the $h_{20}^{\rm (mem, ellip)}$ mode for an elliptical binary described via Eq.~\eqref{eq:hlm-hypergeom}, with $e_{-}=0.99$ and $e_0=0.01$ at a pericenter distance of $r_0\equiv p_0/(1+e_0)=6M$. This mode is plotted as the blue (solid) line in Fig.~\ref{fig:h20-circ-ecc}. We will also need to model how the eccentricity evolves with time. To do this, I evolve Eqs.~\eqref{eq:detdt} and \eqref{eq:drpdt}, but I change to a new time variable $T=-t$ so that I can more easily evolve the system ``backwards'' in time starting from the initial conditions $r_{\rm p}(T=0) \equiv r_0=6M$ and $e_t(T=0) \equiv e_0=0.01$. For an equal-mass binary, I find that the ``early-time'' eccentricity $e_{-}=0.99$ is reached at a time $T_{-}/M\approx 2.031\times 10^8$. This mode is compared with a purely quasicircular model for the $h_{20}^{\rm (mem)}$ mode which is given by
\be
h_{20}^{\rm (mem, circ)}(T) = \frac{2}{7} \sqrt{\frac{10\pi}{3}} \eta \frac{M}{R} \left[ \frac{M}{r(T)} - \frac{M}{r_{-}} \right],
\ee
where
\be
r(T)= r_0 \left(1+\frac{T}{\tau_{\rm rr}}\right)^{1/4}, \;\;\;\text{with}\;\;\; \tau_{\rm rr} = \frac{5}{256} \frac{M}{\eta} \left(\frac{r_0}{M}\right)^4,
\ee
and $r_{-} \equiv r(T_{-})\approx 225.8 M$. This model forces both the quasicircular mode $h_{20}^{\rm (mem, circ)}$ and the elliptical mode $h_{20}^{\rm (mem, ellip)}$ to vanish at the same time ($T=T_{-}$), which can be considered the start of the observation. It also ensures that both orbits have a pericenter separation $r_{\rm p}=r_0=6M$ at time $T=0$. The two modes are plotted in Fig.~\ref{fig:h20-circ-ecc}, where the value of time for both modes is parameterized in terms of the eccentricity of the elliptical mode. This figure indicates that the quasicircular model provides a moderately accurate representation of the true evolution of the memory mode (which accounts for the orbit's past eccentricity). At the end of the evolution ($T=0$, $r_{\rm p}=6M$), the two modes have a fractional error of $\approx 1.5\%$.

To better quantify the degree to which the two modes ``overlap,'' I have computed the following normalized inner product:
\be
{\mathcal O} = \frac{\int_0^{T_{-}} h_1(t) h_2(t) \,dt }{\sqrt{\left[ \int_0^{T_{-}} h_1^2(t) \,dt  \right] \left[ \int_0^{T_{-}} h_2^2(t) \,dt  \right]}},
\ee
where $h_1=h_{20}^{\rm (mem, circ)}$ and $h_2=h_{20}^{\rm (mem, ellip)}$.
This is equivalent to the commonly computed overlap between two GW signals, but here assuming white noise. For values of $e_0=0.01$ or $0.001$, I find the value ${\mathcal O} \approx 0.976$; this decreases slightly to $0.975$ for $e_0=0.1$. Although I have not considered a realistic noise model, this calculation suggests that ignoring the effects of past eccentricity in quasicircular binaries is a reasonable approximation and is not likely to result in significant reduction in the signal-to-noise ratio of the nonlinear memory.

Now let us consider the memory that results over the entire lifetime of a binary system, including its initial formation. As one would intuitively expect, any bound elliptic binary experiencing gravitational radiation-reaction evolves to larger eccentricities (and larger orbital separations) into the past until $e_t>1$ and the binary becomes unbound. This was proved rigorously in \cite{walker-will-I-PRD1979}. Equivalently, for certain choices of its initial orbital parameters, a hyperbolic binary can lose energy from gravitational-wave emission and become bound. The waveform for such a scenario can be approximately modeled using Eqs.~\eqref{eq:d2I20dt2} and \eqref{eq:d2I22dt2} combined with a prescription for the instantaneous evolution of the orbital elements (see, e.g., \cite{walker-will-I-PRD1979}). If we choose our time and angular coordinates conveniently so that capture happens at periastron, a schematic description of the waveform modes from such a captured binary would look like the left-hand plot of Fig.~\ref{fig:Ilm-modes} for $\hat{t}\equiv (t/M)(M/p)^{3/2}<0$ smoothly matched onto the right-hand plot of Fig.~\ref{fig:Ilm-modes} for $\hat{t}>0$. (Note that the different modes and their slopes in that figure have the correct signs at $\hat{t}=0$ to allow for such a matching.) After capture, such a binary would circularize and eventually merge, with the waveforms evolving in the standard way for $\hat{t}>0$. Of course, this description is somewhat idealized. In the real world, other interactions (e.g., tidal dissipation, three-body interactions, gas drag, or dynamical friction) are more likely to result in binary capture (although gravitational radiation losses could play an important role in very dense stellar systems such as globular clusters or galactic nuclei).\footnote{Another possibility is that the binary was not captured but was ``born bound,'' with each component star forming from a fragmenting molecular cloud. The system could then have evolved into a compact-object binary and a source of GWs.} However, for the purpose of considering the size of the memory jump over very long time scales, let us presume that at some early time the binary is in an unbound, hyperbolic orbit, while at some later time it is a bound, elliptic binary that circularizes and merges. For such a binary, the total memory jump is roughly given by Eq.~\eqref{eq:memdiff} with
\bs
\begin{align}
\lim_{t \rightarrow -\infty} h_{+,\times} &\propto \frac{\eta M^2}{Rp_i} \frac{(e_i^2-1)^{3/2}}{e_i^2}, \\
\lim_{t \rightarrow  \infty} h_{+,\times} &\propto \frac{\Delta E}{R},
\end{align}
\es
where $p_i$ and $e_i$ are the semilatus rectum and eccentricity prior to capture\footnote{The eccentricity after capture is approximately given by $e_0=e_i + \Delta e$, where $\Delta e= -\frac{170 \pi}{3} \eta (M/p_i)^{5/2}$. This can be derived by considering the change in eccentricity for a parabolic binary ($e_i=1$),
\be
\Delta e = \int_{-\pi}^{\pi} \frac{de_t}{dv} dv,
\ee
where $de_t/dv=(de_t/dt)/(dv/dt)$. An expression for the instantaneous (i.e., not orbit-averaged) value of $de_t/dt$ can be derived by considering the Lagrange planetary equation (cf.~\cite{danby}) for an osculating Keplerian ellipse under the action of the 2.5PN radiation reaction force (see also the last equation in \cite{walker-will-I-PRD1979} or Eq.~(2.14) of \cite{blanchetschafer1PN}).}, and $\Delta E$ is the energy radiated in GWs throughout the inspiral, merger, and ringdown.\footnote{However, note that the nonlinear memory is only proportional to the radiated energy at leading-order in an $(l,m)$ mode expansion of the energy flux \cite{favata-memory-saturation}.} This suggests that large memory jumps can result not only from the nonlinear memory (which grows most rapidly during the final phases of coalescence), but also through the linear memory associated with binary capture.

A more relevant issue is the observability of some signature of the formation or early-time state of the binary. Clearly, if one's GW detector is operating when the binary capture process occurs (in retarded time), then the signature of the capture, including the resulting memory, will be seen by the detector (provided it is sensitive to low-frequency effects like the memory). However, what if the capture process (and the associated passing GWs) occurred long before the start of the observation period? Does the capture process still leave an ``imprint'' on the waves observed at later times? Intuitively, one expects the answer to this question to be ``no.'' This is indeed correct as can be seen with the following argument.
\begin{figure}[t]
\includegraphics[angle=0, width=0.48\textwidth]{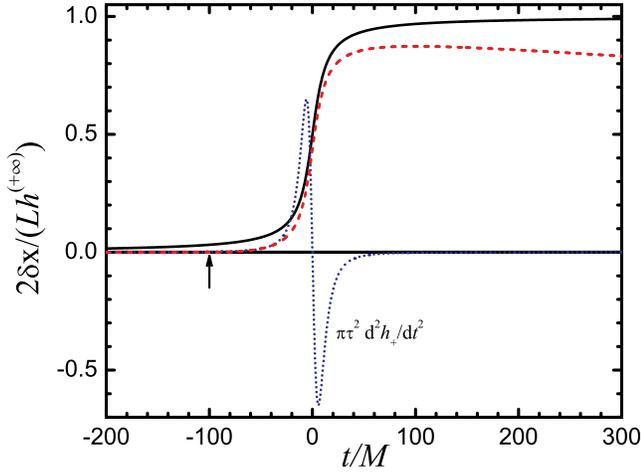}
\caption{\label{fig:deltaxmemory}(color online). Response of a freely falling gravitational-wave detector to a memory signal. The differential displacement [normalized by the detector length $L$ and asymptotic value of the memory $h^{(+\infty)}$] is plotted vs time. The memory signal is modeled by Eq.~\eqref{eq:memarctan} with $h^{(-\infty)}=0$ and $\tau=10 M$. The second time-derivative in Eq.~\eqref{eq:ddmemarctan} is also plotted (blue/dotted curve). The solid (black) curve shows the response of a detector that observes the entire memory signal [cf.~Equation \eqref{eq:deltax1}]. The dashed (red) curve shows the response of a detector that starts monitoring the memory at time $t_0=-10\tau$ (indicated by the arrow) [Eq.\eqref{eq:deltax2}]. In this case the $t<-10\tau$ build-up of the memory is lost, and a linear drift (proportional to the slope at $t_0=-10\tau$) develops at late times.}
\end{figure}

Consider a simplified GW detector consisting of two particles floating in space separated by a distance $L$. Placing the first particle at the origin of its own proper reference frame, the position $x^j$ of the second particle relative to the first is given by the equation of motion (Ch.~35.5 of \cite{mtw})
\be
\ddot{x}^j = \frac{1}{2} \ddot{h}_{jk}^{\rm TT} x^k,
\ee
where overdots here refer to the derivative of the proper time at the first particle, and $h_{jk}^{\rm TT}$ is the metric perturbation in transverse-traceless gauge. We can choose to orient our two particles and the resulting coordinate system such that their motion along their direction of separation $\hat{x}$ is given by
\be
\ddot{x}(t) = \frac{1}{2} \ddot{h}_{+}(t) x(t).
\ee
For very small displacements, $x(t)\approx L+ \delta x(t)$, and the equation for the difference in the particles' relative separation simplifies to
\be
\label{eq:deltax}
\ddot{\delta x}(t) = \frac{L}{2} \ddot{h}_{+}(t).
\ee
Now we consider two scenarios: in the first we assume that our detector has been freely-floating for all times, so it observes the entire build up of the memory. In the distant past, we assume that our memory signal approaches the value $h^{(-\infty)}$ and that its derivative vanishes, $\dot{h}_{+}^{\rm (mem)}(-\infty) \rightarrow 0$. In this case, Eq.~\eqref{eq:deltax} has the solution
\be
\label{eq:deltax1}
\delta x(t) = \frac{L}{2} \left[ h_{+}^{\rm (mem)}(t) -  h^{(-\infty)}  \right].
\ee
Already we see in this case that the value of the GW field in the asymptotic past [$h^{(-\infty)}$] is not observable; instead only the difference between that asymptotic value and the current value (at time $t$) is observable. However, since the detector has been operating for arbitrarily long times, the measured value of the memory retains any imprint of the past evolution of the binary [e.g., the value of the nonlinear memory at time $t$ would depend on the motion of the source all the way to $t\rightarrow -\infty$, but not on the value of $h^{(-\infty)}$].

In the second scenario, let us suppose that the memory signal has started arriving, but our detector is rigidly fixed in position until some  time $t_0$ when we allow our particles to be free-floating. In this case, Eq.~\eqref{eq:deltax} has the solution\footnote{This situation is equivalent to solving Eq.~\eqref{eq:deltax} with the right-hand side multiplied by a Heaviside function $\Theta(t-t_0)$.}
\be
\label{eq:deltax2}
\delta x(t) = \frac{L}{2} \left[ h_{+}^{\rm (mem)}(t) -  h_{+}^{\rm (mem)}(t_0) - \dot{h}_{+}^{\rm (mem)}(t_0) (t-t_0)  \right]\!.
\ee
Here (somewhat obviously) we see that the memory loses its dependence on times before $t_0$. We also see the development of a linear drift (proportional to the slope of the waveform at $t_0$). This drift arises from the initial impulse the detector receives from the passing wave at the moment it is released.

To make the above discussion more explicit, consider a schematic model for a nonlinear memory waveform given by the arctangent function:
\be
\label{eq:memarctan}
h_{+}^{\rm (mem)} = \frac{h^{(+\infty)}-h^{(-\infty)}}{\pi} \arctan\left(\frac{t}{\tau}\right) + \frac{h^{(+\infty)} + h^{(-\infty)}}{2},
\ee
where $h^{(\pm\infty)}$ are the asymptotic values of the memory and $\tau$ is the characteristic rise time of the memory. The second time-derivative of this function is
\be
\label{eq:ddmemarctan}
\ddot{h}_{+}^{\rm (mem)} = -\frac{2 [h^{(+\infty)}-h^{(-\infty)}] t \tau}{\pi (t^2+\tau^2)^2}.
\ee
These functions and the resulting differential displacements are plotted in Fig.~\ref{fig:deltaxmemory} for the two scenarios mentioned above. This graphically illustrates that we can only observe the build up of the memory that occurs while our detector is operating. A similar model could also be based on the hyperbolic tangent function,
\be
\label{eq:memtanh}
h_{+}^{\rm (mem)} = \frac{h^{(+\infty)}-h^{(-\infty)}}{2} \tanh\left(\frac{t}{\tau}\right) + \frac{h^{(+\infty)} + h^{(-\infty)}}{2},
\ee
which approaches its asymptotes more quickly and has a second derivative given by
\be
\ddot{h}_{+}^{\rm (mem)} = -\frac{[h^{(+\infty)}-h^{(-\infty)}] \sinh(t/\tau)}{\tau^2 \cosh^3(t/\tau)}.
\ee
\section{\label{sec:estimates}Estimating the signal-to-noise ratio of memory jumps}
Here I provide some simple formulas for estimating the detectability of the memory. For the case of merging quasicircular binaries, detectability estimates are presented in \cite{favata-memory-saturation} and will be discussed in more detail in \cite{favata-PNNR-memory}. For elliptical binaries, we have seen that the memory behaves quite similarly to the quasicircular case, so the estimates of detectability are little changed. I instead will focus on the linear and nonlinear memory for hyperbolic and parabolic orbits.

We begin by defining the angle-averaged square of the signal-to-noise ratio as
\be
\langle \rho^2 \rangle = \int_0^{\infty} \frac{h_c^2(f)}{h_n^2(f)} \frac{df}{f},
\ee
where the average is over all sky positions, source orientations, and polarization angles [see, e.g., Eqs.~(2.33)\mbox{--}(2.36) of \cite{flanagan-hughesI}].
Here, $h_n(f)=\sqrt{\alpha f S_n(f)}$ is the sky-averaged rms noise amplitude per logarithmic frequency interval. The factor $\alpha$ is $5$ for orthogonal arm detectors like LIGO and $20/3=5/\sin^2(60^{\circ})$ for equilateral triangles like LISA or the Einstein Telescope \cite{barackcutler2}.
The characteristic amplitude is given by\footnote{For cosmological sources, one must replace $f\rightarrow (1+z)f$ and $R\rightarrow D_L(z)/(1+z)$ in this expression, where $z$ is the redshift and $D_L$ is the luminosity distance.}
\be
h_c(f) = 2 f \langle |\tilde{h}_{+}(f)|^2 + |\tilde{h}_{\times}(f)|^2 \rangle^{1/2},
\ee
where a tilde denotes a Fourier transform. If we approximate the memory as a step-function, then its Fourier transform is given by\footnote{This follows from the Fourier transform of the Heaviside function,
\[
\int_{-\infty}^{+\infty} H(\pm t) e^{2\pi i f t} dt = \frac{\delta(f)}{2} \pm \frac{i}{2\pi f}.
\]
This step-function approximation is equivalent to the zero-frequency limit (ZFL) discussed in \cite{turner-neutrinomemory,smarr-zfl,bontz-price,wagoner-lowfreq}. In that case, one approximates the Fourier transform of the time-derivative of a signal $h(t)$ near $f\approx 0$ via
\begin{align}
\tilde{\dot{h}}(f) &= \int_{\infty}^{+\infty} \dot{h}(t) e^{2\pi i f t} dt \approx  \int_{\infty}^{+\infty} \dot{h}(t) dt, \nonumber\\
&=h(+\infty) - h(-\infty) \equiv \Delta h, \nonumber
\end{align}
and we use the usual relation for the Fourier transform of a derivative, $\tilde{\dot{h}}(f) = (-2\pi i f)\tilde{h}(f)$, to arrive at
\[
\tilde{h}_{\rm ZFL}(f) = \frac{i \Delta h}{2\pi f}.
\]
}
\be
|\tilde{h}_{+,\times}(f)|= \frac{\Delta h_{+,\times}}{2\pi f}.
\ee
However, a real memory signal has some finite rise time $\tau$ which imposes a high-frequency cutoff at $f_c \sim 1/\tau$ in the Fourier transform.\footnote{For an explicit example of this, consider the Fourier transform of the signal in Eq.~\eqref{eq:memtanh} for $f>0$ \cite{bracewell}:
\begin{align}
\tilde{h}_{+}^{\rm (mem)} &= \frac{[h^{(+\infty)}-h^{(-\infty)}]}{2} i\pi \tau \csch(\pi^2 \tau f) \nonumber \\
&= \frac{i [h^{(+\infty)}-h^{(-\infty)}]}{2\pi f} \left\{ 1 - \frac{\pi^4}{6} (\tau f)^2 + O[(\tau f)^4] \right\}.\nonumber
\end{align}
Here, one can see from the Taylor expansion the sharp cutoff in the ZFL value of the Fourier transform when $f\sim 1/\tau$.}
We can therefore approximate the characteristic strain by
\be
\label{eq:hcmem}
h_c = \frac{1}{\pi} \langle |\Delta h_{+}|^2 + |\Delta h_{\times}|^2 \rangle^{1/2} \Theta(f_c-f).
\ee
The SNR then becomes
\be
\label{eq:SNR}
\langle \rho^2 \rangle^{1/2} = \frac{\hat{h}_c}{\hat{N}},
\ee
where we define $\hat{h}_c$ to be Eq.~\eqref{eq:hcmem} without the Heaviside factor and
\be
\label{eq:Nhat}
\hat{N} = \left( \int_0^{f_c} \frac{df}{f h_n^2}  \right)^{-1/2}.
\ee
To evaluate $\hat{N}$, one needs to choose a value for the cutoff frequency $f_c \sim \tau^{-1}$. The rise time for a hyperbolic trajectory is $\tau = \kappa M (p/M)^{3/2}$, where $\kappa$ is a factor that depends on how the rise time is defined.\footnote{If we define the rise time by taking the integral in Eq.~\eqref{eq:time-eqn} over $v \in [-\Theta_s/2, \Theta_s/2]$, then we find that $\kappa \leq 4/3$ and asymptotically approaches $2/e_t^3$ for $e_t\gg 1$.} Although $\hat{N}$ clearly depends on the parameters of the binary (through $f_c$), we tabulate its value for several detectors in Table \ref{tab:Nhat} by choosing $f_c$ to be either the location of the minimum of $f h_n^2$ or the high-frequency cutoff for the detector.
\begin{table}[t]
\caption{\label{tab:Nhat}Evaluation of the quantity $\hat{N}$ in Eq.~\eqref{eq:Nhat} for different detectors. The integral is taken from a low-frequency cutoff of $30$ Hz in case of initial LIGO, $10$ Hz for advanced LIGO and advanced Virgo, $1$ Hz for the three iterations of the Einstein Telescope (ET), and $3\times 10^{-6}$ Hz for LISA. In the $\hat{N}_{\rm max}$ column, the upper limit of the integral [$f_c$ in Eq.~\eqref{eq:Nhat}] is taken as $2000$ Hz for the ground-based detectors and $1$ Hz for LISA. In the $\hat{N}_{\rm low}$ column, the upper limit of the integral is taken to be the frequency where $f h_n^2$ is minimized. The sensitivity curves used here can be found in \cite{ligo-noise,advligo-noise,advvirgo-noise,ET-noise,barackcutler}.}
\begin{ruledtabular}
\begin{tabular}{lrr}
Detector & $\hat{N}_{\rm low}$ & $\hat{N}_{\rm max}$ \\
\hline
Initial LIGO & $1.20 \times 10^{-21}$ & $7.20 \times 10^{-22}$	\\
Advanced LIGO & $1.24 \times 10^{-22}$ & $5.61 \times 10^{-23}$ \\
Advanced Virgo & $1.62 \times 10^{-22}$ & $6.93 \times 10^{-23}$ \\
ET-b & $9.35\times 10^{-24}$ & $4.52 \times 10^{-24}$ \\
ET-c & $1.22 \times 10^{-23}$ & $4.61 \times 10^{-24}$ \\
ET-d & $1.74 \times 10^{-23}$ & $5.18 \times 10^{-24}$\\
LISA & $1.70 \times 10^{-21}$ & $7.92 \times 10^{-22}$\\
\end{tabular}
\end{ruledtabular}
\end{table}

For the case of a pulsar timing array (PTA), we can estimate $\hat{N}$ by making use of Eq.~(31) in \cite{pshirkov-etal-memory}:
\be
\hat{N}_{\rm PTA} = 1.94 \times 10^{-16} \left( \frac{250}{N_t} \frac{20}{N_{\alpha}} \right)^{1/2} \left( \frac{10 {\rm yrs}}{T_{\rm obs}} \right) \left( \frac{\sigma_n}{100 {\rm ns}} \right),
\ee
where $N_t$ is the number of measured timing residuals, $N_{\alpha}$ is the number of pulsars in the array, $\sigma_n$ is the noise in the timing residuals (assumed to be Gaussian stationary white noise that is uncorrelated and the same for each pulsar), and $T_{\rm obs}$ is the total observation time for the PTA. The numbers used above are for near-future PTAs. The Square Kilometre Array (SKA) \cite{skaweb} will achieve better sensitivity, including a factor of 10 decrease in the noise, and perhaps a factor of $\sim 100$ increase in the number of suitable pulsars \cite{smits-etal-SKApulsardistance-AA2011}.

Taking the angle-averages of Eqs.~\eqref{eq:Deltahpx-linear}, \eqref{eq:Deltahpx-parabolic}, and \eqref{eq:Deltahpx-hyperbolic}, and expressing the results in terms of the pericenter distance, yields the following characteristic amplitudes for parabolic and hyperbolic orbits:
\bs
\label{eq:angleavgh}
\begin{align}
\hat{h}_{c, e_t=0}^{\rm (mem)} &= \frac{\sqrt{110\,100\,291}}{4032} \eta^2 \frac{M}{R} \left( \frac{M}{r_{\rm p}} \right)^{7/2}, \\
\hat{h}_{c, e_t\gg1}^{\rm (lin.~mem)} &= \frac{8}{\pi} \sqrt{\frac{2}{5}} \eta \frac{M}{R} \frac{M}{r_{\rm p}}, \\
\hat{h}_{c, e_t\gg1}^{\rm (mem)} &= \frac{\sqrt{21\,075\,910}}{3600} \eta^2 \frac{M}{R} \left( \frac{M}{r_{\rm p}} \right)^3 V_{\infty},
\end{align}
\es
where the second and third equations show the linear and nonlinear memory for hyperbolic orbits in the large-eccentricity limit (for which the memory is largest). Note that in the hyperbolic case, the nonlinear memory is smaller than the linear memory by a factor $\approx 0.79 \eta (M/r_p)^2 V_{\infty}$. In practice, this amounts to a factor $\gg 4$ orders of magnitude, so we ignore the nonlinear memory in the hyperbolic case. Plugging in numbers for some plausible (but perhaps optimistic) sources yields
\bs
\label{eq:hestimate}
\begin{align}
\hat{h}_{c, e_t=0}^{\rm (mem)} &= 2.2\times 10^{-22} \left( \frac{\eta}{0.25} \right)^2 \left( \frac{M/10 M_{\odot}}{R/10 \text{kpc}}\right) \! \left( \frac{20M}{r_p} \right)^{7/2} \!\!, \nonumber \\
&= 1.7 \times 10^{-29} \left( \frac{\eta}{10^{-5}} \right)^2 \! \left( \frac{M/10^6 M_{\odot}}{R/20 \text{Mpc}}\right) \!\! \left( \frac{20M}{r_p} \right)^{7/2} \!\!\!, \nonumber  \\
&= 2.2 \times 10^{-22} \left( \frac{\eta}{0.25} \right)^2 \left( \frac{M/10^6 M_{\odot}}{R/1 \text{Gpc}}\right) \! \left( \frac{20M}{r_p} \right)^{7/2} \!\!\!,
\end{align}
\begin{align}
\hat{h}_{c, e_t\gg1}^{\rm (lin.~mem)} &= 9.6\times 10^{-19} \left( \frac{\eta}{0.25} \right) \left( \frac{M/10 M_{\odot}}{R/10 \text{kpc}}\right) \! \left( \frac{20M}{r_p} \right) \!,  \nonumber  \\
&= 1.9\times 10^{-21} \left( \frac{\eta}{10^{-5}} \right) \! \left( \frac{M/10^6 M_{\odot}}{R/20 \text{Mpc}}\right) \!\! \left( \frac{20M}{r_p} \right) \!, \nonumber  \\
&= 9.6 \times 10^{-19} \left( \frac{\eta}{0.25} \right) \left( \frac{M/10^6 M_{\odot}}{R/1 \text{Gpc}}\right) \! \left( \frac{20M}{r_p} \right)\!.
\end{align}
\es
[For cosmological distances we should take $M/R \rightarrow M_z/D_L$ in the above expressions, where $M_z=(1+z)M$ is the redshifted mass.] The SNR can then be estimated by combining Eq.~\eqref{eq:hestimate} with Eq.~\eqref{eq:SNR} and the numbers in Table \ref{tab:Nhat}. These rough estimates indicate that GW bursts with linear memory could be detectable with second-generation ground-based detectors and future space-based detectors. The nonlinear memory from GW bursts from unbound (or marginally) bound binaries will be more difficult to detect and will likely require third-generation detectors. Current and near-term PTAs are not sufficiently sensitive to detect memory bursts of the types considered here.
\section{\label{sec:conclusion}Conclusions}
This work has generalized previous computations of the nonlinear gravitational-wave memory effect to the case of binaries with arbitrary eccentricity. In the case of hyperbolic, parabolic, and radial orbits, the nonlinear memory is a 2.5PN correction to the waveform. In the case of elliptical binaries, the nonlinear memory affects the waveform at leading order (just as in the quasicircular case). To completely describe elliptical waveforms at leading-PN-order, the nonlinear memory contributions derived here should be added to the well-known nonmemory expressions first derived by Peters and Mathews \cite{petersmathews} and Wahlquist \cite{wahlquist}.
I have also investigated the sensitivity of the nonlinear memory to the early-time history of the binary. In the case of quasicircular binaries that were initially elliptical, the early-time eccentricity provides only a small correction to the memory. Furthermore, I have shown that contributions to the memory made outside of the observation time are undetectable. Lastly, I provided simple estimates of the signal-to-noise ratio for memory bursts arising from sources on unbound orbits.

There are a variety of areas in which this study could be extended. The nonlinear-memory calculations presented here are restricted to leading order. For hyperbolic, parabolic, and radial orbits the waveforms are only known to 1PN order, so there is little motivation to compute higher-order corrections to the leading-order nonlinear memory terms (which themselves enter as 2.5PN-order corrections to the waveform). However, in the elliptical case, the oscillatory waveform polarizations are known to 2PN order, so 2PN-order corrections to the leading-order nonlinear memory terms would be needed to have complete 2PN-order elliptic waveforms. In addition, the effects of spinning binary components on the nonlinear memory have not yet been computed. This calculation is in progress in the case of quasicircular binaries and will be reported elsewhere \cite{favata-spinningmemory}. Computing the nonlinear memory for eccentric, spinning binaries will be left for future work.  It would also be interesting to investigate the size of the linear and nonlinear memory in the case of ultrarelativistic collisions and scatterings \cite{death-highspeed-PRD1978,death-payne-I,death-payne-II,death-payne-III,sperhake-etal-highEcoll-PRL2008,sperhake-etal-highEcoll-PRL2009}. These situations could show a very large memory effect.
\begin{acknowledgments}
This research was supported through an appointment to the NASA Postdoctoral Program at the Jet Propulsion Laboratory, administered by Oak Ridge Associated Universities through a contract with NASA. Early phases of this work were also supported by the National Science Foundation under grant no.~PHY05-51164 to the Kavli Institute for Theoretical Physics. I am grateful to Yanbei Chen for useful discussions and to K.~G.~Arun, Curt Cutler, Xinyi Guo, and Bala Iyer for their helpful comments on this manuscript.
\end{acknowledgments}
\appendix
\section{\label{app:moments}
DERIVATION OF THE LEADING-ORDER MASS AND CURRENT SOURCE MOMENTS FOR A GENERAL TWO-BODY SYSTEM}
The purpose of this appendix is to derive expressions for the mass and current multipole moments in the form of $(l,m)$ modes that are valid for any two-body orbit at Newtonian order. At leading order, we are only concerned about the so-called \emph{source moments} which are defined in terms of integrals over a stress-energy pseudotensor. General expressions (valid for any PN order) for the mass and current symmetric-trace-free (STF) source multipoles, ${\mathcal I}_L$ and ${\mathcal J}_L$, can be found in Eq.~(85) of \cite{blanchetLRR}. These STF tensors with $L$ indices (where $L\equiv a_1 a_2 \cdots a_l$) can be difficult to work with, and for some calculations it is more convenient to instead use the ``scalarized'' versions of these moments, $I_{lm}$ and $J_{lm}$. These ``scalar'' multipoles are simply the coefficients of the expansion of the STF mass and current multipoles on the basis of the STF spherical harmonics ${\mathcal Y}_L^{lm}$ [these are defined in Eq.~(2.12) of \cite{kiprmp} and are related to the standard scalar spherical harmonics via Eq.~\eqref{eq:Ylm} below]. The STF moments and their $(l,m)$ modes are related by the following formulas [Eqs.~(4.6) and (4.7) of \cite{kiprmp}]:
\bs
\label{eq:IJstfdef}
\begin{align}
\label{eq:Istfdef}
{\mathcal I}_L &= \frac{l!}{4} \sqrt{\frac{2 l (l-1)}{(l+1)(l+2)}} \sum_{m=-l}^{l} I_{lm} {\mathcal Y}^{lm}_L , \\
\label{eq:Jstfdef}
{\mathcal J}_L &= -\frac{(l+1)!}{8l} \sqrt{\frac{2 l (l-1)}{(l+1)(l+2)}} \sum_{m=-l}^{l} J_{lm} {\mathcal Y}^{lm}_L \;,
\end{align}
\es
\bs
\label{eq:IJlmdef}
\begin{align}
\label{eq:Ilmdef}
\text{and} \;\;\; I_{lm} &= A_l {\mathcal I}_L {\mathcal Y}^{lm \, \ast}_L , \\
\label{eq:Jlmdef}
J_{lm} &= B_l {\mathcal J}_L {\mathcal Y}^{lm \, \ast}_L, \;\; \text{where}
\end{align}
\es
\bs
\label{eq:ABl}
\begin{align}
\label{eq:Al}
A_l &=\frac{16 \pi}{(2l+1)!!} \sqrt{\frac{(l+1)(l+2)}{2l(l-1)}}, \\
\label{eq:Bl}
B_l &=-\frac{32 \pi l}{(2l+1)!!} \sqrt{\frac{(l+2)}{2l(l+1)(l-1)}}.
\end{align}
\es

Now we specialize the general form for STF mass and current moments in Eq.~(85) of \cite{blanchetLRR}  to the Newtonian-order moments for general orbits (but arbitrary $l$-value). This derivation could be easily extended to the 1PN-order moments (see \cite{kidder08,damour-iyer-nagar}).
The Newtonian-order source mass and current multipole moments for a system of $N$ (nonspinning) point masses is
\bs
\label{eq:IJLNbodies}
\begin{align}
\label{eq:ILNbody}
{\mathcal I}_L^{{\rm N}} &= \sum_{A=1}^N m_A y_A^{\langle L \rangle} , \\
\label{eq:JLNbody}
{\mathcal J}_L^{{\rm N}} &= \sum_{A=1}^N m_A \varepsilon^{a b \langle i_l} y_A^{L-1 \rangle} y_A^a v_A^b ,
\end{align}
\es
where $A$ labels the body, the multi-index $L$ refers to a product of $l$ vectors (e.g., $y_1^L=y_1^{a_1} y_1^{a_2} \cdots y_1^{a_l}$), $\varepsilon^{abc}$ is the Levi-Civita tensor, and the angled brackets $\langle \rangle$ mean to take the STF projection on the enclosed indices. The ``${\rm N}$'' superscript emphasizes that these are Newtonian-order moments.
We now specialize to a two-body system with masses $m_1$ and $m_2$, total mass $M=m_1+m_2$, and reduced mass ratio $\eta=m_1 m_2/M^2$. We transform to the center-of-mass frame using
\bs
\label{eq:ycom}
\begin{align}
\label{eq:y1com}
\vec{y}_1 &= \frac{m_2}{M} \vec{x} ,\\
\label{eq:y2com}
\vec{y}_2 &= -\frac{m_1}{M} \vec{x} ,
\end{align}
\es
where $\vec{x} = \vec{y}_1 - \vec{y}_2 = r\vec{n}$ has length $r$ and $\vec{n}$ points from $m_2$ to $m_1$. We also define the individual and relative velocity vectors via $\vec{v}_A=\dot{\vec{y}}_A$ and $\vec{v}=\dot{\vec{x}}$.

Substituting the above relations into Eqs.~\eqref{eq:IJLNbodies} gives [Eqs.~(5.21) and (5.22) of \cite{blanchet3pnwaveform}]:
\bs
\label{eq:IJLN}
\begin{align}
\label{eq:ILN}
{\mathcal I}_L^{{\rm N}} &= \eta M s_l(\eta) x_{\langle L \rangle},  \\
\label{eq:JLN}
{\mathcal J}_L^{{\rm N}} &= \eta M s_{l+1}(\eta) \varepsilon_{a b \langle i_l} x_{L-1 \rangle} x^a v^b \,, \;\;\; \text{where}
\end{align}
\es
\be
\label{eq:sl}
s_l(\eta) = X_2^{l-1} + (-1)^{l} X_1^{l-1}  \,,
\ee
and we define $X_1 \equiv \frac{m_1}{M} = \frac{1}{2}(1+\Delta)$, $X_2 \equiv \frac{m_2}{M} = \frac{1}{2}(1-\Delta)$, and $\Delta \equiv \frac{m_1-m_2}{M} = \pm \sqrt{1-4\eta}$ (the $\pm$ sign depends on one's convention for which mass is larger).

To compute the ``scalar'' multipoles defined in Eq.~\eqref{eq:IJlmdef} we need to contract Eqs.~\eqref{eq:IJLN} with ${\mathcal Y}_L^{lm \, \ast}$. Using the relationship between the ``scalar'' and STF spherical harmonics,
\be
\label{eq:Ylm}
Y^{lm} = {\mathcal Y}_L^{lm} n_L = {\mathcal Y}_L^{lm} n_{\langle L \rangle},
\ee
the Newtonian ``scalar'' mass multipole equivalent to \eqref{eq:ILN} is easily seen to be
\be
\label{eq:IlmN}
I_{lm}^{{\rm N}} = A_l \eta M s_l(\eta) r^l Y^{lm \ast}(\theta,\phi) \;.
\ee
To derive the Newtonian ``scalar'' current multipole moment we use the definition of the magnetic-type ``pure-spin'' vector spherical harmonics [Eqs.~(2.18b) and (2.23b) of \cite{kiprmp}]:
\begin{align}
\label{eq:YBlm}
Y^{B,lm}_b &= \sqrt{\frac{l}{l+1}} \varepsilon_{b a i_l} n_a {\mathcal Y}^{lm}_{i_l L-1} n_{L-1}, \\
& = \frac{1}{\sqrt{l(l+1)}} \varepsilon_{bcd} x_c \nabla_d Y_{lm}.
\end{align}
Combining this equation with Eqs.~\eqref{eq:Jlmdef} and \eqref{eq:JLN} yields the Newtonian scalar current multipole for general orbits:
\be
\label{eq:JlmN}
J_{lm}^{{\rm N}} = \frac{B_l}{l} \eta M s_{l+1}(\eta) r^l (\vec{x} \times \vec{v}) \cdot \vec{\nabla}Y_{lm}^{\ast}(\theta,\phi) \;.
\ee
In Eqs.~\eqref{eq:IlmN} and \eqref{eq:JlmN}, the moments are given as functions of time by solving the equations of motion to determine the spherical coordinates of the relative separation vector $\vec{x}$: $r(t)$, $\theta(t)$, and $\phi \equiv \varphi(t)$. If we restrict ourselves to orbits in the $x$\mbox{--}$y$ plane, we can further simplify the multipole moments by using
\begin{align}
\vec{x} \times \vec{v} &= r^2 \dot{\phi} \vec{e}_z \,, \;\;\; \text{and} \\
\vec{e}_z \cdot \vec{\nabla}Y_{lm} &= -\frac{\sin \theta}{r} \frac{\partial Y_{lm}}{\partial \theta} \;.
\end{align}
The resulting Newtonian-order ``scalar'' multipole moments for general orbits restricted to the $x$\mbox{--}$y$ plane are
\bs
\label{eq:IJxyplane}
\begin{align}
\label{eq:Ilmxy}
I_{lm}^{{\rm N}} &= A_l \eta M s_l(\eta) r^l Y_{lm}^{\ast}\left(\frac{\pi}{2},\phi \right) , \;\;\; \text{and} \\
\label{eq:Jlmxy}
J_{lm}^{{\rm N}} &= -\frac{B_l}{l} \eta M s_{l+1}(\eta) r^{l+1} \dot{\phi} \left. \frac{\partial Y_{lm}^{\ast}(\theta,\phi)}{\partial \theta} \right|_{\theta=\frac{\pi}{2}} .
\end{align}
\es
\section{\label{app:hypergeom}
NONLINEAR MEMORY INTEGRAL FOR ELLIPTICAL ORBITS IN TERMS OF HYPERGEOMETRIC FUNCTIONS}
In this appendix, we show how to derive explicit expressions for the integrals of Eqs.~\eqref{eq:dhlmdet}, which we rewrite as:
\begin{widetext}
\bs
\label{eq:dhlmdet-app}
\begin{align}
h_{20}^{\rm (mem)} &= \left[ -\frac{384}{7 (304)^b} \sqrt{\frac{10\pi}{3}} \frac{\eta M^2}{R p_0} \right] C_0(e_0)   \left[ \int de_t \frac{(1+\frac{145}{48}e_t^2 +\frac{73}{192}e_t^4)}{e_t^{31/19}(1+\frac{121}{304}e_t^2)^{b}} \right] + K_{20} ,\\
h_{2\pm2}^{\rm (mem)} &= \left[ -\frac{52}{7}\frac{\sqrt{5\pi}}{(304)^b} \frac{\eta M^2}{R p_0} \right] C_0(e_0) e^{\mp 2 i \varpi} \left[ \int de_t \frac{e_t^{7/19}(1+\frac{2}{13} e_t^2)}{(1+\frac{121}{304}e_t^2)^{b}} \right] + K_{2\pm2} ,\\
h_{40}^{\rm (mem)} &= \left[ -\frac{64}{21 (304)^b} \sqrt{\frac{\pi}{10}} \frac{\eta M^2}{R p_0} \right] C_0(e_0)   \left[ \int de_t \frac{(1+\frac{99}{32}e_t^2 +\frac{51}{128}e_t^4)}{e_t^{31/19}(1+\frac{121}{304}e_t^2)^{b}} \right] + K_{40} ,\\
h_{4\pm2}^{\rm (mem)} &= \left[ -\frac{13}{21}\frac{\sqrt{\pi}}{(304)^b} \frac{\eta M^2}{R p_0} \right] C_0(e_0) e^{\mp 2 i \varpi} \left[ \int de_t \frac{e_t^{7/19}(1+\frac{2}{13} e_t^2)}{(1+\frac{121}{304}e_t^2)^{b}} \right] + K_{4\pm2} ,\\
h_{4\pm4}^{\rm (mem)} &= \left[ \frac{25}{24 (304)^b} \sqrt{\frac{\pi}{7}} \frac{\eta M^2}{R p_0} \right] C_0(e_0) e^{\mp 4 i \varpi} \left[ \int de_t \frac{e_t^{45/19}}{(1+\frac{121}{304}e_t^2)^{b}} \right] + K_{4\pm4} ,
\end{align}
\es
where $b\equiv \frac{3169}{2299}$, the $K_{lm}$ are integration constants, and we refer to the constants in square brackets as $A_{lm}$ below and in the main text. We now note that all of the indefinite integrals in square brackets can be expressed in terms of combinations of the following integral \cite{mathematica}:
\be
\label{eq:hypergeomint}
\int \frac{x^a}{(1+cx)^b} dx = \frac{x^{a+1}}{a+1} {}_{2}F_1(b,a+1;a+2;-cx),
\ee
where ${}_{2}F_1$ is the hypergeometric function \cite{abramowitz-stegun}. Any of the integrals in Eqs.~\eqref{eq:dhlmdet-app} can then be computed from:
\begin{multline}
\label{eq:Flm}
F_{lm}(e_t) \equiv \int de_t \frac{e_t^{\alpha_{lm}} (1+c_{lm} e_t^2 + d_{lm} e_t^4)}{(1+\beta e_t^2)^b} = e_t^{\alpha_{lm}} \left[ \frac{1}{\alpha_{lm}+1} {}_{2}F_1(b,\tfrac{\alpha_{lm}+1}{2};\tfrac{\alpha_{lm}+3}{2};-\beta e_t^2)
\right. \\ \left.
+ \frac{c_{lm} e_t^2}{\alpha_{lm}+3} {}_{2}F_1(b,\tfrac{\alpha_{lm}+3}{2};\tfrac{\alpha_{lm}+5}{2};-\beta e_t^2) + \frac{d_{lm} e_t^4}{\alpha_{lm}+5} {}_{2}F_1(b,\tfrac{\alpha_{lm}+5}{2};\tfrac{\alpha_{lm}+7}{2};-\beta e_t^2) \right],
\end{multline}
\end{widetext}
where the constants $\alpha_{lm}$, $c_{lm}$, and $d_{lm}$ are easily read off of Eqs.~\eqref{eq:dhlmdet-app}, and $\beta\equiv \frac{121}{304}$.
The integration constants $K_{lm}$ are then determined by the requirement that the nonlinear memory vanish at early times when the eccentricity $e_t=e_{-}$. The final result for the $h_{lm}^{\rm (mem)}$ modes is then given by
\be
h_{lm}^{\rm (mem)} = A_{lm} C_0(e_0) e^{\mp i m \varpi} \left[ F_{lm}(e_t) - F_{lm}(e_{-}) \right].
\ee
\section{\label{app:averaging}
THE ROLE OF AVERAGING THE GRAVITATIONAL-WAVE STRESS-ENERGY TENSOR IN NONLINEAR MEMORY CALCULATIONS}
\begin{figure}[t]
\includegraphics[angle=0, width=0.48\textwidth]{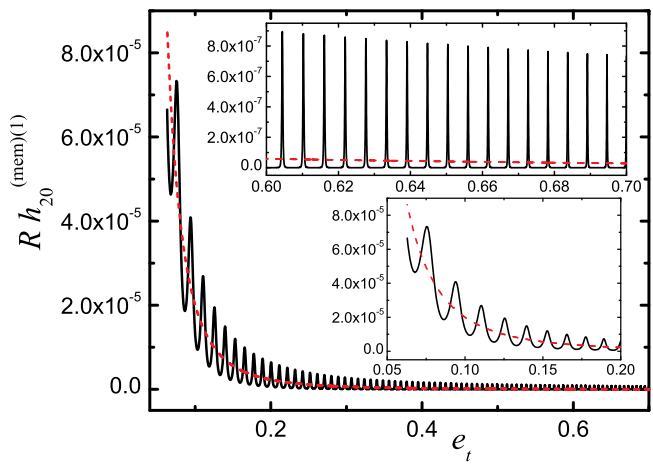}
\caption{\label{fig:et-dh20-compare}(color online). Effect of wavelength averaging on the nonlinear memory integrand. The solid (black) curve shows the time evolution (parameterized in terms of decreasing $e_t$) of the integrand $h_{20}^{\rm (mem)(1)}$ computed without wavelength averaging via Eq.~\eqref{eq:dh20dt-kep} as described in the text. The dashed (red) curve shows $h_{20}^{\rm (mem)(1)}$ computed with wavelength averaging via Eq.~\eqref{eq:dh20dt-ecc}. The averaging procedure removes the short time scale oscillations from the integrand. The insets zoom in on the low and high eccentricity regions.}
\end{figure}
\begin{figure*}[t]
$
\begin{array}{cc}
\includegraphics[angle=0, width=0.5\textwidth]{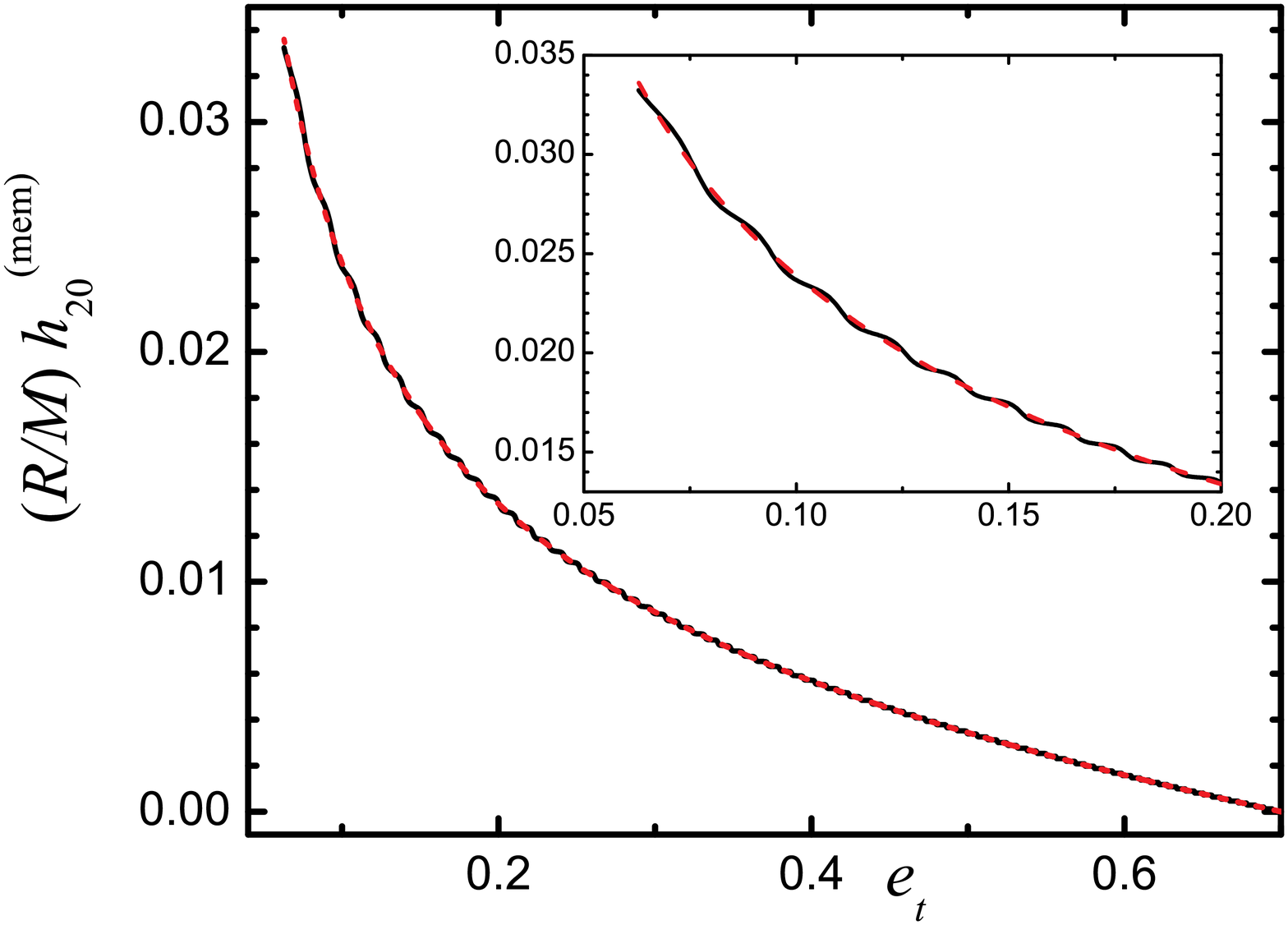} &
\includegraphics[angle=0, width=0.48\textwidth]{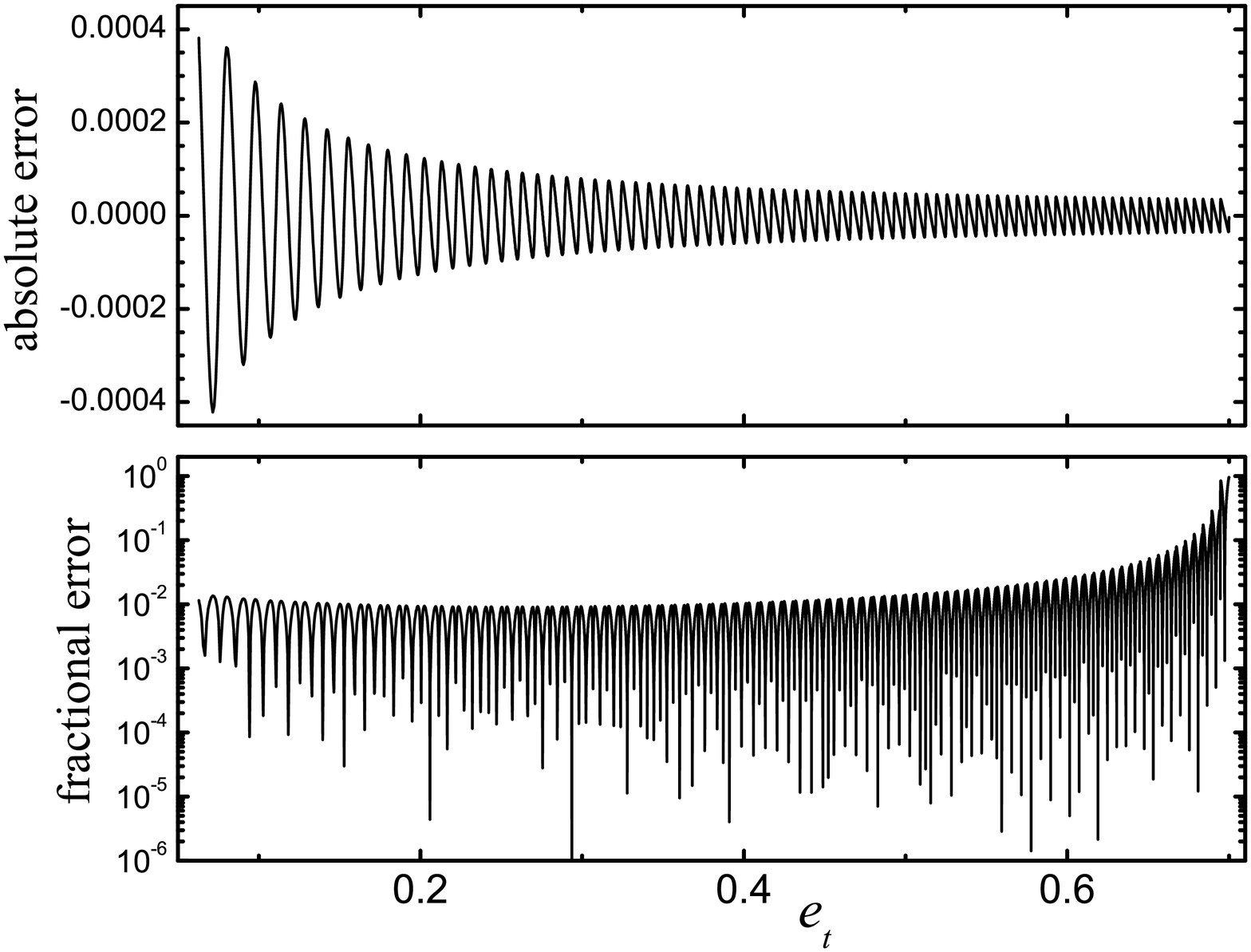}
\end{array}
$
\caption{\label{fig:et-h20-compare}(color online). Effect of wavelength averaging on the nonlinear memory mode. The left-hand plot shows the $h_{20}^{\rm (mem)}$ nonlinear memory mode (the time integral of Fig.~\ref{fig:et-dh20-compare}). The solid (black) curve is without wavelength averaging; the dashed (red) curve is with averaging. The curves lie nearly on top of each other, aside from small amplitude oscillations that remain when one integrates the nonaveraged integrand. The inset zooms in on the low-eccentricity region. The integration is terminated at the last-stable-orbit. The right-hand plots show the absolute and fractional errors between the two curves.}
\end{figure*}
In the definition of the nonlinear memory in Eq.~\eqref{eq:hlmmem}, an explicit averaging over several wavelengths appears in the gravitational-wave energy flux $\frac{dE_{\rm gw}}{dt d\Omega}$. This is consistent with the standard derivation in which averaging is necessary to obtain a well-defined GW stress-energy tensor \cite{isaacson,mtw}. However, in the derivations of the nonlinear memory in \cite{christodoulou-mem,blanchet-damour-hereditary}, this wavelength averaging does not explicitly appear. The purpose of this appendix is to investigate (in the context of eccentric binaries) how the nonlinear memory calculation depends on whether the wavelength averaging is performed or not. The short answer to this question is that the nonlinear memory does not depend on this averaging, aside from very small amplitude oscillations at the orbital period that are superimposed on the memory when averaging is \emph{not} performed. The reason why the memory is relatively insensitive to the averaging procedure is simple: performing the time integral that explicitly appears in the nonlinear memory calculation effectively ``averages'' over the integrand [cf.~Equation\eqref{eq:hlmmem}]. So by performing also the wavelength averaging $\langle \rangle$ of the integrand, one is effectively ``averaging'' twice. However, note that the wavelength averaging significantly simplifies the integrand, allowing for an analytic calculation.\footnote{Note also that in the circular, nonspinning case, this wavelength averaging issue does not arise. The integrand of those modes that contribute to the nonlinear memory are constant on an orbital time scale and are unaffected by averaging. This can be seen explicitly by comparing Eqs.~\eqref{eq:dhmem-kep} and \eqref{eq:dhlmdt-ecc} in the $e_t=0$ case. We are currently investigating this averaging issue for the case of quasicircular, spinning binaries \cite{favata-spinningmemory}.}

To investigate this issue in more detail, we can explicitly compute the $h_{20}^{\rm (mem)}$ nonlinear memory mode with and without wavelength averaging. In both cases, we first solve for the evolution of $e_t(t)$, $p(t)$, and $v(t)$ (the true anomaly) by numerically integrating Eqs.~\eqref{eq:dpdt-dedt} and \eqref{eq:dphidt}. We assume an equal-mass binary and initial conditions $e_t(0)=0.7$, $p(0)=30 M$, and $v(0)=0$. We then substitute the result into the integrand of the time integral for the memory: Eq.~\eqref{eq:dh20dt-kep} in the nonaveraged case (ignoring the $\langle \rangle$), and Eq.~\eqref{eq:dh20dt-ecc} in the averaged case. The resulting integrands are plotted in Fig.~\ref{fig:et-dh20-compare}. There we see that if we do not perform any wavelength averaging, the integrand retains oscillations at the orbital period; these oscillations are smoothed over by the wavelength averaging.

We then numerically integrate both the averaged and nonaveraged integrands, starting from the condition $h_{20}^{\rm mem}(0)=0$. The result is plotted in the left-hand side of Fig.~\ref{fig:et-h20-compare}. There we see that performing the wavelength averaging has had very little effect on the resulting memory. The two curves lie nearly on top of each other. As stated above, this agreement is simply due to the fact that the time integration of the nonaveraged $h_{20}^{\rm (mem)(1)}$ essentially acts as an averaging procedure. The only difference is a very small remnant oscillation about the wavelength-averaged curve. At the last-stable-orbit and for a large range of eccentricity, the two curves agree to $\sim 1\%$.
\bibliography{eccentricmem-textV2}
\end{document}